\documentclass[aps,prd,twocolumn,superscriptaddress,amssymb,eqsecnum,showpacs,showkeyes,secnumarabic,graphics,floatfix,nofootinbib,tightenlines,longbibliography]{revtex4-1}
\usepackage{graphicx}
\usepackage{bm}
\usepackage{cancel}
\usepackage{dcolumn}
\usepackage{amsmath}
\usepackage{hhline}
\usepackage[utf8]{inputenc}
\usepackage[dvipsnames]{xcolor}
\usepackage[breaklinks=true,colorlinks=true,
linkcolor=blue,urlcolor=Blue,citecolor=MidnightBlue,
bookmarks=true,bookmarksopenlevel=2]{hyperref}

\def \beq  {\begin{equation}}
\def \eeq  {\end{equation}}
\def \ber  {\begin{eqnarray}}
\def \eer  {\end{eqnarray}}
\def \Geff {G_{\rm eff}}

\def \omms   {\Omega_{\rm m}}

\def \Geff {G_{\rm eff}}

\begin{document}
\newcommand{\newc}{\newcommand}

\newc{\be}{\begin{equation}}
\newc{\ee}{\end{equation}}
\newc{\ba}{\begin{eqnarray}}
\newc{\ea}{\end{eqnarray}}
\newc{\bea}{\begin{eqnarray*}}
\newc{\eea}{\end{eqnarray*}}
\newc{\D}{\partial}
\newc{\ie}{{\it i.e.} }
\newc{\eg}{{\it e.g.} }
\newc{\etc}{{\it etc.} }
\newc{\etal}{{\it et al.}}
\newc{\lcdm}{$\Lambda$CDM }
\newc{\omom}{$\Omega_{0m}$ }
\newc{\plcdm}{Planck15/$\Lambda$CDM }
\newcommand{\fs}{{\rm{\it f\sigma}}_8}

\newcommand{\nn}{\nonumber}
\newc{\ra}{\Rightarrow}
\title{Tension and constraints on modified gravity parametrizations of $G_{\textrm{eff}}(z)$ from growth rate and Planck data}

\author{Savvas Nesseris}\email{savvas.nesseris@csic.es}
\affiliation{Instituto de F\'isica Te\'orica UAM-CSIC, Universidad Auton\'oma de Madrid, Cantoblanco, 28049 Madrid, Spain}
\author{George Pantazis}\email{gpantaz@cc.uoi.gr}
\affiliation{Department of Physics, University of Ioannina, GR-45110, Ioannina, Greece}
\author{Leandros Perivolaropoulos}\email{leandros@uoi.gr}
\affiliation{Department of Physics, University of Patras, GR-26500, Patras, Greece\\ (on leave from Department of Physics, University of Ioannina, GR-45110, Ioannina, Greece)}

\date{\today}

\begin{abstract}
We construct an updated and extended compilation of growth rate data based on recent Redshift Space Distortion (RSD) measurements. The dataset consists of 34 datapoints and includes corrections for model dependence. In order to minimize overlap and maximize the independence of the datapoints we also construct a subsample of this compilation (a `Gold' growth dataset) which consists of 18 datapoints. We test the consistency of this dataset with the best fit \plcdm parameters in the context of General Relativity (GR) using the evolution equation for the growth factor $\delta(a)$ with a $w$CDM background. We find tension at the $\sim 3 \sigma$ level between the best fit parameters $w$ (the dark energy equation of state), $\Omega_{0m}$ (the matter density parameter) and $\sigma_8$ (the matter power spectrum normalization on scales $8h^{-1}$\textrm{Mpc}) and the corresponding \plcdm  parameters ($w =-1$, $\Omega_{0m} = 0.315$ and $\sigma_8 = 0.831$). We show that the tension disappears if we allow for evolution of the effective Newton constant, parametrized as $G_{\textrm{eff}}(a)/G_{\textrm{N}} = 1 + g_a(1-a)^n-g_a(1-a)^{2n}$ with $n\ge2$ where $g_a$ and $n$ are parameters of the model, $a$ is the scale factor and $z = 1/a-1$ is the redshift. This parametrization satisfies three important criteria: a) positive energy of graviton ($G_{\textrm{eff}} > 0$), b) consistency with Big Bang Nucleosynthesis constraints ($G_{\textrm{eff}}(a\ll1)/G_{\textrm{N}}=1$) and c) consistency with Solar System tests ($G_{\textrm{eff}}(a=1)/G_{\textrm{N}}=1$ and $G_{\textrm{eff}}'(a=1)/G_{\textrm{N}}=0$). We show that the best fit form of $G_{\textrm{eff}}(z)$ obtained from the growth data corresponds to weakening gravity at recent redshifts (decreasing function of $z$) and we demonstrate that this behavior is not consistent with any scalar-tensor Lagrangian with a real scalar field. Finally, we use MGCAMB to find the best fit $G_{\textrm{eff}}(z)$ obtained from the Planck CMB power spectrum on large angular scales and show that it is a mildly increasing function of $z$, in $3\sigma$ tension with the corresponding decreasing best fit $G_{\textrm{eff}}(z)$ obtained from the growth data.
\end{abstract}
\maketitle

\section{Introduction}
\label{sec:Introduction}

Despite the vast improvement in quality and quantity of the cosmological observations during the past 18 years, the simplest cosmological model predicting an accelerating expansion of the Universe, known as the \lcdm \cite{Bull:2015stt-lcdm-review}, has remained viable and consistent with observations \citep{Tsujikawa:2010sc-dark-energy-review,Caldwell:2009ix-dark-energy-review,Copeland:2006wr-review-dark-en}. Crucial assumptions of this model are the validity of General Relativity (GR) on cosmological scales, flatness homogeneity, isotropy and the invariance of dark energy in both space and time (cosmological constant). The parameters of this model have been pinned down to extraordinary accuracy by the Planck \citep{Ade:2015xua-planck-2015-params} mission.  These parameter values define the concordance \plcdm model and are shown in Table~\ref{tab:planck}. This model is consistent with a wide range of independent cosmological observations testing mainly the large scale cosmological background $H(z)$. Such observations include earlier analyses of cosmic microwave background (CMB) fluctuations \cite{Hinshaw:2012aka-wmap-cosmo-params},  large scale velocity flows \cite{Watkins:2014zaa-velocity-flows}, baryon acoustic oscillations \citep{Aubourg:2014yra-bao-data,Delubac:2014aqe-bao-data1}
Type Ia supernovae \cite{Betoule:2014frx-jla-snia-data}, early growth rate of perturbations data\cite{Huterer:2013xky-growth-data1,Nesseris:2015fqa-growth-data2,Basilakos:2012uu-growth-data2,Nesseris:2007pa-growth-data3}, gamma ray burst data \cite{Izzo:2015vya-grb-dark-energy,Wei:2013xx-grb-dark-energy1,Samushia:2009ib-grb-dark-energy1},
strong and weak lensing data \cite{Baxter:2016ziy-weak-lensing}, $H(z)$ (Hubble parameter) data \cite{Ding:2015vpa-lcdm-tension-hofz-data}, HII galaxy data  \cite{Chavez:2016epc-hii-galaxies}, cluster gas mass fraction data
\cite{Allen:2007ue-gas-mass-fraction,Morandi:2016cet-gas-mass-fraction2}.

Despite of the consistency of \plcdm with large cosmological scales background data, it has become evident recently that a mild tension appears to exist between \plcdm and some independent observations in intermediate cosmological scales ($z\leq 0.6$),  \cite{Raveri:2015maa-tension-review}. Such tensions include estimates of the Hubble parameter \cite{Bernal:2016gxb-h0-tension2,Roukema:2016wny-h0-tension1, Abdalla:2014cla-bao-anomaly2,Meng:2015loa-h0-tension2,Qing-Guo:2016ykt-lcdm-h0-tension2,Sola:2016jky,Sola:2017jbl}
in the context of \lcdm, estimates of  the amplitude of the power spectrum on the scale of $8 h^{-1} \textrm{Mpc}$  ($\sigma_8$) \cite{Bull:2015stt-lcdm-review} and estimates of the matter density parameter $\Omega_{0m}$ \cite{Gao:2013pfa-om0-tension}.

In addition, there are theoretical arguments based on naturalness that may hint toward physics beyond the concordance \lcdm model \citep{Tsujikawa:2010sc-dark-energy-review,Caldwell:2009ix-dark-energy-review,Copeland:2006wr-review-dark-en}.

The data that are in some tension with \plcdm appear to indicate consistently that there is a lack of gravitational power in structures on intermediate-small cosmological scales. This lack of power may be expressed through different cosmological parameters in a degenerate manner. For example, it may be expressed as a lower value of \omom at redshifts less than about $0.6$ or as a smaller value of $\sigma_8$ or as a dark energy equation of state that becomes smaller than $-1$ at low redshifts.

The situation is reminiscent of the corresponding situation in the early 90's before the confirmation of \lcdm by SnIa data \cite{Perlmutter:1998np-lcdm-conf,Riess:1998cb-lcdm-conf} when the Einstein-de Sitter flat ``standard CDM model" was seen to be in mild tension with a range of cosmological data on large cosmological data including the COBE discovery of large scale CMB fluctuations which were larger than expected in the CDM model. It was first realized by Efstathiou in 1990 that there is more power on large scales than predicted by CDM \cite{Efstathiou:1990xe-first-lcdm} and that a flat universe with a cosmological constant could ease the large scale tension. This analysis was confirmed by other subsequent studies\cite{Moscardini:1993nw-first-lcdm,Gorski:1991rk-first-lcdm,Ratra:1994dm-first-lcdm,Kofman:1993ag-first-lcdm,Krauss:1995yb}. Despite the evidence that the CDM model lacked the required power on large large scales to match observations, it remained the ``standard model" until 1998 when the accelerating expansion was confirmed at several $\sigma$ using Type Ia supernovae \cite{Perlmutter:1998np-lcdm-conf,Riess:1998cb-lcdm-conf}.

The parameter that is most commonly used to describe the lack of power of \plcdm on small scales is the variance of the linear matter perturbations on $8h^{-1}\textrm{Mpc}$,  $\sigma_8$. This parameter can be obtained from a weak lensing correlation function obtained by the CFHTLenS collaboration \cite{Heymans:2013fya-CFHTLenS-low-s8}, from the galaxy cluster count\cite{Ade:2013lmv-planck13-galaxy-counts-low-s8} and from Redshift Space Distortion (RSD) data\cite{Macaulay:2013swa-growth-tension-s8,Alam:2016hwk-dmr12-complete-fs8dat}. These datasets indicate that there is lower growth power than the one inferred in the context of \plcdm and GR, at about $2\sigma$ level \cite{MacCrann:2014wfa-s8-tension,Battye:2014qga-s8-tension,Kitching:2016hvn-s8-tension}. This tension if not due to systematics, could be reconciled by a mechanism that reduces the rate of clustering between recombination and today. Three such possible mechanisms are as follows:
\begin{itemize}
\item
A Hot Dark Matter component induced \eg by a sterile neutrino\cite{Hamann:2013iba-sterile-hdm-component}
\item
Dark matter clusters differently at small and large scales, a possibility explored in Ref. \cite{Kunz:2015oqa}.
\item
Modifications of GR\cite{Mueller:2016kpu-growth-modif-grav} which attenuate the growth rate of perturbations.
\end{itemize}
In the present study we focus on the third mechanism. If a modification of GR is responsible for the observed cosmological accelerating expansion it would also lead to a modified growth rate of cosmological density perturbations compared to the one predicted in GR. This growth rate has been measured in several surveys in redshifts ranging from $z=0.02$ up to $z=1.4$ and is defined as
\be
f(a)=\frac{d\delta(a)}{d\ln a}
\label{gfdef}
\ee
where $\delta(a)\equiv \frac{\delta\rho}{\rho}$ denotes the cosmological overdensity and $a(t)$ is the scale factor.

Most growth rate measurements are obtained using peculiar velocities obtained from RSD measurements \cite{Kaiser:1987qv} identified in galaxy redshift surveys.
In general such surveys can provide measurements of the perturbations in terms of the galaxy density $\delta_g$, which are related to matter perturbations through the bias parameter $b$ as $\delta_g=b\; \delta_m$. Thus early growth rate measurements provided values of the growth rate $f$ divided by the bias factor $b$ leading to the parameter $\beta\frac{f}{b}$.

This measured parameter is sensitive to the value of the bias $b$ which can vary in the range $b\in [1,3]$. This uncertainty factor makes it difficult to combine values of $\beta$ from different regions and different surveys leading to unreliable datasets of $\beta(z_i)$.

A more reliable combination is the product $f(z)\sigma_8(z)\equiv f\sigma_8(z)$, as it is independent of the bias and may be obtained using either weak lensing or RSD. Thus, in the present study we only consider surveys that have reported the growth rate in the robust form $f(z)\sigma_8(z)$. These surveys along with the corresponding datapoints are shown in Table~\ref{tab:fs8-data-all} where the data are shown in chronological order, along with the assumed fiducial cosmology and other notes, \eg their covariance matrix and so on.

Some of these points are in fact highly correlated with other points since they were produced by analyses of the same sample of galaxies. Also, it is clear from Table~\ref{tab:fs8-data-all} that there has been a dramatic increase and improvement of the growth rate data during the past five years. This is mainly due to the SDSS, BOSS, WiggleZ, and Vipers surveys that have dramatically increased the number of growth rate data and their constraining power. The quality and quantity of the growth rate data are expected to improve dramatically in the coming years with the Euclid \cite{Amendola:2016saw} and LSST \cite{Abell:2009aa} surveys.

Despite the dramatic improvement of the quality and quantity of the growth rate data their combination into a single uniform and self-consistent dataset remains a challenge. There are two basic reasons for this:
\begin{itemize}
\item
{\bf Model Dependence:} Since surveys do not measure distances to galaxies directly, they have to assume a specific cosmological model in order to infer distances. All growth rate datapoints shown in Table~\ref{tab:fs8-data-all} assume a flat \lcdm cosmological background albeit with different \omom and/or $\sigma_8$. The actual values of these parameters used for each datapoint are shown in Table~\ref{tab:fs8-data-all}. This model dependence requires a correction before the data are included in a single uniform dataset.
\item
{\bf Double Counting:} Some of the data points shown in Table~\ref{tab:fs8-data-all} correspond to the same sample of galaxies analyzed by different groups/methods and the inclusion of all these points without proper corrections would lead to double-counting and artificial decrease of the error regions.
\end{itemize}
In the present analysis we address the above issues and construct a new large, uniform and reliable growth rate dataset which consists of independent datapoints that are corrected for model dependence by rescaling growth rate measurements by proper ratios of $H(z) D_A(z)$ where $D_A(z)$ is the angular diameter distance. We use this dataset to investigate the tension level with a \plcdm background model under the assumption of validity of GR.

The tension we find can be eliminated by either changing the background Hubble parameter $H(z)$ or by allowing modifications of GR through a scale independent effective Newton constant $G_{\text{eff}}(z)$. We follow that latter route, and assuming that the \plcdm background is correct, we find the best fit form of $G_{\text{eff}}(z)$ using the \plcdm $H(z)$ and our growth rate dataset. The derivation of the best fit effective Newton's constant along with the \plcdm $H(z)$ allows the reconstruction of the underlying fundamental model Lagrangian density in the context of specific classes of models.

A general and generic such class of models is scalar-tensor theories where the action in the Jordan frame is determined by the scalar field potential $U(\phi)$ and the nonminimal coupling $F(\phi)$ in the form
{\small{
\be
{\cal S}= \int d^4 x \sqrt{-g} \left[ \frac{1}{2}F(\phi)R - \frac{1}{2}Z(\phi) g^{\mu\nu} \partial_\mu \phi \partial_\nu \phi - U(\phi) \right] + S_m,
\label{action}
\ee }}

\begin{table}[t!]
\begin{centering}
\caption{\plcdm parameters with $68\%$ limits. Based on TT,TE,EE+lowP and a flat $\Lambda$CDM model (middle column) or a $w$CDM model (right column); see Table 4 of Ref. \citep{Ade:2015xua-planck-2015-params} and the Planck chains archive \footnote{A pdf describing the data contained in the Planck archive can be found here \url{<https://wiki.cosmos.esa.int/planckpla2015/index.php/Cosmological_Parameters>}}.
\label{tab:planck}}
\begin{tabular}{|ccc|}
  \hhline{===}
  Parameter & Value ($\Lambda$CDM) & Value ($w$CDM) \\
    \hhline{===}
$\Omega_b h^2$ & $0.02225\pm0.00016$ & $0.02229\pm0.00016$ \\
$\Omega_c h^2$ & $0.1198\pm0.0015$ & $0.1196\pm0.0015$\\
$n_s$ & $0.9645\pm0.0049$ & $0.9649\pm0.0048$\\
$H_0$ & $67.27\pm0.66$ & $>81.3$\\
$\Omega_m$ & $0.3156\pm0.0091$ & $0.203^{+0.022}_{-0.065}$\\
$w$ & $-1$ & $-1.55^{+0.19}_{-0.38}$\\
$\sigma_8$ & $0.831\pm0.013$ & $0.983^{+0.100}_{-0.055}$\\
  \hhline{===}
\end{tabular}
\end{centering}
\end{table}

\noindent where $R$ is the Ricci scalar. We have set $8\pi G_{\textrm{N}} \equiv 1$ for simplicity (and therefore $F_0 = 1$ at the present time) and $S_m$ is the matter action of some arbitrary matter fields, \ie does not involve the scalar field $\phi$. Even though the scalar field is fully described by the set of $F(\phi), Z(\phi)$ and $U(\phi)$, a convenient reduction to two parameters can be applied (\eg Refs. \cite{Nesseris:2006jc, EspositoFarese:2000ij, Perrotta:1999am}) by a rescaling of the scalar field. For example, we may have the Brans-Dicke reduction where $F(\phi)=\phi, \ Z(\phi)=\omega(\phi)/\phi$, or alternatively we can obtain $Z(\phi)=1$ with arbitrary $F(\phi)$ as done in the present analysis. We note that all of the above are applied in the Jordan frame where the model is studied. In addition, $F(\phi)>0$ is required so that gravitons have positive energy and $dF/d\phi< 4 \times 10^{-4}$ according to Solar System tests (see Refs.~ \cite{Will:2005va,EspositoFarese:2000ij}). As discussed in section \ref{sec:Section 4}, the effective Newton constant $G_\text{eff}(z)$ is approximately  inversely proportional to the nonminimal coupling $F(\phi(z))$ and is observable through the growth of cosmological perturbations.

It is thus possible to use the best fit form of $G_\text{eff}(z)$ along with the \plcdm $H(z)$ to reconstruct the underlying scalar-tensor theory potential $U(\phi)$ that would produce the observed functional forms of $G_\text{eff}(z)$ and $H(z)$. This scalar-tensor theory is defined by the functional forms of the scalar field potential $U(\phi)$ and nonminimal coupling $F(\phi)$ that are reconstructed uniquely using the method of Refs \cite{Boisseau:2000pr,EspositoFarese:2000ij,Nesseris:2006jc}. However, as also noted in Ref.~\cite{EspositoFarese:2000ij}, this task is not always possible as the reconstructed kinetic term of the scalar field in many cases becomes negative at some redshift range $z$, \ie $\phi^{\prime}(z)^2 <0$, and as a result, the field itself becomes imaginary. In what follows, we derive the properties of functions $F(z)$ that lead to positive kinetic terms for a real scalar field when used in a reconstruction. These properties come from the fact that $F(z)$ satisfies a differential inequality and we can deduce them by using the Chaplygin theorem on differential inequalities.

The structure of this paper is the following. In the next section we introduce the new robust and extended growth dataset (Table~\ref{tab:fs8-data-gold}) and use it to investigate the tension level between growth data and \plcdm in the context of GR. In section \ref{sec:Section 3} we allow for extensions of GR and introduce $G_{\text{eff}}(z)$ parametrizations consistent with Solar System tests and nucleosynthesis. We then find the best fit form of $G_{\text{eff}}(z)$ for each parametrization and investigate the effect of the evolving Newton's constant on the tension between growth data and \plcdm. In section \ref{sec:Section 4} we use the best fit forms of $G_{\text{eff}}(z)$ to implement the reconstruction method for the derivation of the underlying scalar-tensor potential. We find that for the particular form of the best fit $G_\text{eff}(z)$ no consistent reconstruction of a realistic scalar-tensor model can be implemented due to the fact that the kinetic term of the scalar field becomes negative, \ie $\phi^{\prime}(z)^2<0$. Then, by using the Chaplygin theorem on differential inequalities, we derive the required properties of the observed $G_{\textrm{eff}}(z)$ in the context of a \lcdm background so that a well defined scalar-tensor theory can be reconstructed. In section \ref{sec:isw} we determine the effects of the $G_{\textrm{eff}}(z)$ parametrization on the low-$\ell$ multipoles of the CMB, while in section \ref{sec:Conclusion} we conclude, summarize and  discuss future extensions of the present work.

\section{Extended Calibrated Growth RSD dataset: Tension with \plcdm}
\label{sec:Section 2}
\subsection{Theoretical Background}

\begin{table*}[t!]
\caption{A collection of recent $f\sigma_8(z)$ measurements from different surveys, ordered chronologically. In the columns we show the name and year of the survey that made the measurement, the redshift and value of $f\sigma_8(z)$ and the corresponding reference and fiducial cosmology. These datapoints are not independent and should not be used all together at the same time. For a robust compilation, see Table \ref{tab:fs8-data-gold}.
\label{tab:fs8-data-all}}
\begin{centering}
\begin{tabular}{|ccccccc|}
\hhline{=======}
Index & Dataset & $z$ & $f\sigma_8(z)$ & Refs. & Year & Notes \\
\hline
1& SDSS-LRG & $0.35$ & $0.440\pm 0.050$ & \cite{Song:2008qt} & 2006 &$(\Omega_m,\Omega_K)=(0.25,0)$ \\

2& VVDS & $0.77$ & $0.490\pm 0.18$ & \cite{Song:2008qt}  & 2008 & $(\Omega_m,\Omega_K)=(0.25,0)$ \\

3 & 2dFGRS & $0.17$ & $0.510\pm 0.060$ & \cite{Song:2008qt}  & 2009& $(\Omega_m,\Omega_K)=(0.3,0)$ \\

4 & 2MASS &0.02& $0.314 \pm 0.048$ &  \cite{Davis:2010sw},\cite{Hudson:2012gt} & 2010& $(\Omega_m,\Omega_K)=(0.266,0)$ \\

5 & SnIa+IRAS &0.02& $0.398 \pm 0.065$ &  \cite{Turnbull:2011ty},\cite{Hudson:2012gt} & 2011& $(\Omega_m,\Omega_K)=(0.3,0)$\\

6 &SDSS-LRG-200 & $0.25$ & $0.3512\pm 0.0583$ & \cite{Samushia:2011cs} & 2011& $(\Omega_m,\Omega_K)=(0.25,0)$  \\

7 &SDSS-LRG-200 & $0.37$ & $0.4602\pm 0.0378$ & \cite{Samushia:2011cs} & 2011& \\

8&SDSS-LRG-60 & $0.25$ & $0.3665\pm0.0601$ & \cite{Samushia:2011cs} &2011& $(\Omega_m,\Omega_K)=(0.25,0)$ \\

9&SDSS-LRG-60 & $0.37$ & $0.4031\pm0.0586$ & \cite{Samushia:2011cs} & 2011&\\

10 &WiggleZ & $0.44$ & $0.413\pm 0.080$ & \cite{Blake:2012pj} & 2012&$(\Omega_m,h)=(0.27,0.71)$ \\

11 &WiggleZ & $0.60$ & $0.390\pm 0.063$ & \cite{Blake:2012pj} & 2012&$C_{ij}\rightarrow$ Eq.~(\ref{wigglez}). \\

12 &WiggleZ & $0.73$ & $0.437\pm 0.072$ & \cite{Blake:2012pj} & 2012 &\\

13&SDSS-BOSS& $0.30$ & $0.407\pm 0.055$ & \cite{Tojeiro:2012rp} & 2012 & $(\Omega_m,\Omega_K)=(0.25,0)$ \\

14&SDSS-BOSS& $0.40$ & $0.419\pm 0.041$ & \cite{Tojeiro:2012rp} & 2012 & \\

15&SDSS-BOSS& $0.50$ & $0.427\pm 0.043$ & \cite{Tojeiro:2012rp} & 2012 & \\

16&SDSS-BOSS& $0.60$ & $0.433\pm 0.067$ & \cite{Tojeiro:2012rp} & 2012 & \\

17& SDSS-DR7-LRG & $0.35$ & $0.429\pm 0.089$ & \cite{Chuang:2012qt}  &2012 &$(\Omega_m,\Omega_K)=(0.25,0)$\\

18&6dFGRS& $0.067$ & $0.423\pm 0.055$ & \cite{Beutler:2012px} & 2012&$(\Omega_m,\Omega_K)=(0.27,0)$ \\

19& GAMA & $0.18$ & $0.360\pm 0.090$ & \cite{Blake:2013nif}  & 2013& $(\Omega_m,\Omega_K)=(0.27,0)$ \\

20& GAMA & $0.38$ & $0.440\pm 0.060$ & \cite{Blake:2013nif}  & 2013& \\

21&BOSS-LOWZ& $0.32$ & $0.384\pm 0.095$ & \cite{Sanchez:2013tga}  &2013 & $(\Omega_m,\Omega_K)=(0.274,0)$ \\

22& SDSS-CMASS & $0.59$ & $0.488\pm 0.060$ & \cite{Chuang:2013wga} &2013& $\ \ (\Omega_m,h,\sigma_8)=(0.307115,0.6777,0.8288)$ \\

23&Vipers& $0.80$ & $0.470\pm 0.080$ & \cite{delaTorre:2013rpa} &2013& $(\Omega_m,\Omega_K)=(0.25,0)$  \\

24& SDSS-MGS & $0.15$ & $0.490\pm0.145$ & \cite{Howlett:2014opa} & 2014& $(\Omega_m,h,\sigma_8)=(0.31,0.67,0.83)$ \\

25& SDSS-veloc & $0.10$ & $0.370\pm 0.130$ & \cite{Feix:2015dla}  &2015 &$(\Omega_m,\Omega_K)=(0.3,0)$ \\

26&FastSound& $1.40$ & $0.482\pm 0.116$ & \cite{Okumura:2015lvp}  & 2015& $(\Omega_m,\Omega_K)=(0.270,0)$\\

27& 6dFGS+SnIa & $0.02$ & $0.428\pm 0.0465$ & \cite{Huterer:2016uyq} & 2016 & $(\Omega_m,h,\sigma_8)=(0.3,0.683,0.8)$ \\

28&Vipers PDR-2& $0.60$ & $0.550\pm 0.120$ & \cite{Pezzotta:2016gbo} & 2016& $(\Omega_m,\Omega_b)=(0.3,0.045)$ \\

29&Vipers PDR-2& $0.86$ & $0.400\pm 0.110$ & \cite{Pezzotta:2016gbo} & 2016&\\

30& BOSS DR12 & $0.38$ & $0.497\pm 0.045$ & \cite{Alam:2016hwk} &2016& $(\Omega_m,\Omega_K)=(0.31,0)$ \\

31& BOSS DR12 & $0.51$ & $0.458\pm 0.038$ & \cite{Alam:2016hwk} &2016& \\

32& BOSS DR12 & $0.61$ & $0.436\pm 0.034$ & \cite{Alam:2016hwk} &2016& \\

33&Vipers v7& $0.76$ & $0.440\pm 0.040$ & \cite{Wilson:2016ggz} &2016& $(\Omega_m,\sigma_8)=(0.308,0.8149)$ \\

34&Vipers v7 & $1.05$ & $0.280\pm 0.080$ & \cite{Wilson:2016ggz} & 2016&\\
\hhline{=======}
\end{tabular}\par\end{centering}
\end{table*}

\begin{table*}[t!]
\caption{A compilation of robust and independent $f\sigma_8(z)$ measurements from different surveys, based on Table \ref{tab:fs8-data-all}. In the columns, we show in ascending order with respect to redshift, the name and year of the survey that made the measurement, the redshift and value of $f\sigma_8(z)$, and the corresponding reference and fiducial cosmology. These datapoints are used in our analysis in the next sections.
\label{tab:fs8-data-gold}}
\begin{centering}
\begin{tabular}{|ccccccc|}
\hhline{=======}
Index & Dataset & $z$ & $f\sigma_8(z)$ & Refs. & Year & Notes \\
\hline
1 & 6dFGS+SnIa & $0.02$ & $0.428\pm 0.0465$ & \cite{Huterer:2016uyq} & 2016 & $(\Omega_m,h,\sigma_8)=(0.3,0.683,0.8)$ \\

2 & SnIa+IRAS &0.02& $0.398 \pm 0.065$ &  \cite{Turnbull:2011ty},\cite{Hudson:2012gt} & 2011& $(\Omega_m,\Omega_K)=(0.3,0)$\\

3 & 2MASS &0.02& $0.314 \pm 0.048$ &  \cite{Davis:2010sw},\cite{Hudson:2012gt} & 2010& $(\Omega_m,\Omega_K)=(0.266,0)$ \\

4 & SDSS-veloc & $0.10$ & $0.370\pm 0.130$ & \cite{Feix:2015dla}  &2015 &$(\Omega_m,\Omega_K)=(0.3,0)$ \\

5 & SDSS-MGS & $0.15$ & $0.490\pm0.145$ & \cite{Howlett:2014opa} & 2014& $(\Omega_m,h,\sigma_8)=(0.31,0.67,0.83)$ \\

6 & 2dFGRS & $0.17$ & $0.510\pm 0.060$ & \cite{Song:2008qt}  & 2009& $(\Omega_m,\Omega_K)=(0.3,0)$ \\

7 & GAMA & $0.18$ & $0.360\pm 0.090$ & \cite{Blake:2013nif}  & 2013& $(\Omega_m,\Omega_K)=(0.27,0)$ \\

8 & GAMA & $0.38$ & $0.440\pm 0.060$ & \cite{Blake:2013nif}  & 2013& \\

9 &SDSS-LRG-200 & $0.25$ & $0.3512\pm 0.0583$ & \cite{Samushia:2011cs} & 2011& $(\Omega_m,\Omega_K)=(0.25,0)$  \\

10 &SDSS-LRG-200 & $0.37$ & $0.4602\pm 0.0378$ & \cite{Samushia:2011cs} & 2011& \\

11 &BOSS-LOWZ& $0.32$ & $0.384\pm 0.095$ & \cite{Sanchez:2013tga}  &2013 & $(\Omega_m,\Omega_K)=(0.274,0)$ \\

12 & SDSS-CMASS & $0.59$ & $0.488\pm 0.060$ & \cite{Chuang:2013wga} &2013& $\ \ (\Omega_m,h,\sigma_8)=(0.307115,0.6777,0.8288)$ \\

13 &WiggleZ & $0.44$ & $0.413\pm 0.080$ & \cite{Blake:2012pj} & 2012&$(\Omega_m,h)=(0.27,0.71)$ \\

14 &WiggleZ & $0.60$ & $0.390\pm 0.063$ & \cite{Blake:2012pj} & 2012&$C_{ij}\rightarrow$ Eq.~(\ref{wigglez}). \\

15 &WiggleZ & $0.73$ & $0.437\pm 0.072$ & \cite{Blake:2012pj} & 2012 &\\

16 &Vipers PDR-2& $0.60$ & $0.550\pm 0.120$ & \cite{Pezzotta:2016gbo} & 2016& $(\Omega_m,\Omega_b)=(0.3,0.045)$ \\

17 &Vipers PDR-2& $0.86$ & $0.400\pm 0.110$ & \cite{Pezzotta:2016gbo} & 2016&\\

18 &FastSound& $1.40$ & $0.482\pm 0.116$ & \cite{Okumura:2015lvp}  & 2015& $(\Omega_m,\Omega_K)=(0.270,0)$\\
\hhline{=======}
\end{tabular}\par\end{centering}
\end{table*}

In order to discriminate between GR and modified gravity theories we need an extra observational probe which can track the dynamical properties of gravity. One such probe is the growth function of the linear matter density contrast $\delta\equiv\frac{\delta\rho_m}{\rho_m}$, where $\rho_m$ represents the background matter density and $\delta\rho_m$ represents its first order perturbation.

It can be shown that in many classes of modified gravity theories the growth factor $\delta(a)$ satisfies the following equation \cite{DeFelice:2010gb, Tsujikawa:2007gd, DeFelice:2010aj, Nesseris:2009jf}:
\be
\delta''(a)+\left(\frac{3}{a}+\frac{H'(a)}{H(a)}\right)\delta'(a)
-\frac{3}{2}\frac{\omms \Geff(a,k)/G_{\textrm{N}}}{a^5 H(a)^2/H_0^2}~\delta(a)=0,
\label{eq:ode}
\ee
where primes denote differentiation with respect to the scale factor, $H(a)\equiv\frac{\dot{a}}{a}$ is the Hubble parameter, and $G_{\textrm{eff}}(a,k)$ is the effective Newton constant which is constant and equal to $G_{\textrm{N}}$ in GR. In modified gravity theories, $G_\text{eff}$ depends on both the scale factor $a$ (or equivalently the redshift $z$) and the scale $k$. However, $G_{\textrm{eff}}$ is independent of the scale $k$ for scales smaller than the horizon  ($k\gg a H$) \cite{Sanchez:2010ng}. Thus, on subhorizon scales, we may ignore the dependence on the scale $k$ for both $\delta$ and $G_{\textrm{eff}}$.

For the growing mode we assume the initial conditions $\delta(a\ll1)=a$ and $\delta'(a\ll1)=1$, where in practice we will choose a small enough value of the scale factor so that we are well within the matter domination era, \eg $a_{ini}\sim 10^{-3}$. Note that this equation is only valid on subhorizon scales, \ie $k^2\gg a^2H^2$, where $k$ is the wave-number of the modes of the perturbations in Fourier space. The effects of modified gravity theories enter Eq.~(\ref{eq:ode}) via both $H(a)$ and $\Geff(a,k)$. This is due to the fact that the growth of the large scale structure is a result of the motion of matter and therefore is sensitive to both the expansion of the Universe and the evolution of Newton's ``constant".

In the case of GR, the exact solution of Eq.~(\ref{eq:ode}) for a flat model with a constant dark energy equation of state $w$  is given for the growing mode by \cite{Silveira:1994yq,Percival:2005vm}:
\ba
\delta(a)= a \cdot {}_2F_1 \left(- \frac{1}{3 w},\frac{1}{2} -
\frac{1}{2 w};1 - \frac{5}{6 w};a^{-3 w}(1 - \omms^{-1})\right),\nn\\
\label{Da1}
\ea
where ${}_2F_1(a,b;c;z)$ is a hypergeometric function defined by the series
\be {}_2F_1(a,b;c;z)\equiv \frac{\Gamma(c)}{\Gamma(a)\Gamma(b)}\sum^{\infty}_{n=0}\frac{\Gamma(a+n)\Gamma(b+n)}{\Gamma(c+n)n!}z^n \ee on the disk $|z|<1$ and by analytic continuation elsewhere (see Ref.~\cite{handbook} for more details). In general, it is impossible to find analytical solutions to Eq.~(\ref{eq:ode}) for a generic modified gravity model, so numerical methods for solving it have to be used.

As discussed in the Introduction, a robust measurable quantity in redshift surveys is not the growth factor $\delta(a)$. Instead, it is the combination
\ba
\fs(a)&\equiv& f(a)\cdot \sigma(a)\nn\\
&=&\frac{\sigma_8}{\delta(1)}~a~\delta'(a) ,\label{eq:fs8}
\ea
where $f(a)=\frac{d ln\delta}{d lna}$ is the growth rate and $\sigma(a)=\sigma_8\frac{\delta(a)}{\delta(1)}$  is  the redshift-dependent rms fluctuations of the linear density field within spheres of radius $R=8 h^{-1} \textrm{\textrm{Mpc}}$, while the parameter $\sigma_8$ is its value today. This combination is used in what follows to derive constraints for theoretical model parameters

\subsection{RSD Measurements}

Redshift-space distortions are very important probes of large scale structure providing measurements of $\fs(a)$. This can be achieved by measuring the ratio of the monopole and the quadrupole multipoles of the redshift-space power spectrum which depends on $\beta=f/b$, where $f$ is the growth rate and $b$ is the bias, in a specific way defined by linear theory \cite{Percival:2008sh,Song:2008qt,Nesseris:2006er}. The combination of $f\sigma_8(a)$ is independent of bias  
as all bias dependence in this combination cancels out
thus, it has been shown that this combination is be a good  
discriminator of DE (Dark energy) models \cite{Song:2008qt}.

In Table \ref{tab:fs8-data-all}, we present a collection of recent $f\sigma_8(z)$ measurements from different surveys, ordered chronologically. In the columns, we show the name and year of the survey that made the measurement, the redshift, the value of $f\sigma_8(z)$ and the corresponding reference and fiducial cosmology. The information in some of these datapoints overlaps significantly with other datapoints in the same Table. Some of them are updates on previous measurements either with enhancements in the volume of the survey, during its scheduled run or with different methodologies by various groups. Therefore, the collection of these datapoints should not be used in its entirety.

We thus construct the ``Gold-2017" compilation of robust and independent $f\sigma_8(z)$ measurements from different surveys, shown in Table \ref{tab:fs8-data-gold}. In the columns of Table \ref{tab:fs8-data-gold}, we show the name and year of the survey that made the measurement, the redshift, and value of $f\sigma_8(z)$, and the corresponding reference and fiducial cosmology. These datapoints are used in our analysis in the next sections. These points are a subset of those from Table \ref{tab:fs8-data-all} and were chosen so that only the latest version or more robust version of a measurement is included from every corresponding survey.

In both Tables \ref{tab:fs8-data-all} and \ref{tab:fs8-data-gold}  the data have a dependence on the fiducial model used by the collaborations to convert redshifts to distances, an important step in the derivations of the data. This can be corrected by either taking into account how the correlation function $\xi(r)$ transforms by changing the cosmology, an approach followed by Ref.~\cite{Alam:2015rsa}, or by simply rescaling the growth-rate measurements by the ratios of $H(z)D_A(z)$ of the cosmology used to that of the fiducial one as in Ref.~ \cite{Macaulay:2013swa}. As noted in Ref. \cite{Macaulay:2013swa}, the correction itself is quite small, so we follow the latter method for simplicity.

Specifically, we implement the correction as follows. First, we define the ratio of the product of the Hubble parameter $H(z)$ and the angular diameter distance $d_A(z)$ for the model at hand to that of the fiducial cosmology, \ie
\be
\textrm{ratio}(z)= \frac{H(z)d_A(z)}{H^{fid}(z)d_A^{fid}(z)}, \label{ratio}
\ee
where the values of the fiducial cosmology, namely $\Omega_{0m}$, are given in Table \ref{tab:fs8-data-gold}. Note that the combination $H(z)d_A(z)$ does not depend on $H_0$, so it could be equivalently written in terms of the dimensionless Hubble parameter $E(z)=H(z)/H_0$ and angular diameter distance $D_A(z)=\frac{H_0}{c}d_A(z)$.

Having done this, we can now define the $\chi^2$ as usual for correlated data. We can define a vector $V^i(z_i,p^j)$, where $z_i$ is the redshift of $i$th point and $p^j$ is the $j$th component of a vector containing the cosmological parameters $(\Omega_{0m},w,\sigma_8 \dots)$ that we want to determine from the data. This vector contains the differences of the data and the theoretical model, after we implement our correction. Specifically, it is given by
\be
V^i(z_i,p^j)=f\sigma_{8,i}-\textrm{ratio}(z_i) f\sigma_{8}(z_i,p^j)
\ee
where $f\sigma_{8,i}$ is the value of the $i$th datapoint, with $i=1, \dots, N$, where $N$ is the total number of points, while $f\sigma_{8}(z_i,p^j)$ is the theoretical prediction, both at redshift $z_i$.

Then, the $\chi^2$ can be written as
\be
\chi^2_{growth}=V^i~C_{ij}^{-1}~V^j, \label{chi2growth}
\ee
where $C_{ij}^{-1}$ is the inverse covariance matrix of the data and for compactness we only used the superscripts $i$ and $j$ for the data vectors. For an approximation we will assume that most of the data are not correlated, with the exception of the ones from Wigglez, where the covariance matrix is given by \cite{Blake:2012pj}

\be
C_{ij}^{\text{WiggleZ}}=10^{-3}\left(
         \begin{array}{ccc}
           6.400 & 2.570 & 0.000 \\
           2.570 & 3.969 & 2.540 \\
           0.000 & 2.540 & 5.184 \\
         \end{array}
       \right). \label{wigglez}
\ee
Therefore, the total covariance matrix will be the identity $N\times N$ matrix, but with the addition of a $3\times 3$ matrix at the position of the WiggleZ data, \ie schematically, we could write it as
\be
C_{ij}^{\textrm{growth,total}}=\left(
         \begin{array}{cccc}
           \sigma_1^2 & 0 & 0 & \cdots \\
           0 & C_{ij}^{WiggleZ} & 0& \cdots \\
           0 & 0 & \cdots & \sigma_N^2 \\
         \end{array}
       \right). \label{totalcij}
\ee

An alternative approach would be that of Ref.~\cite{Alam:2015rsa} where the authors approximated the total covariance matrix of all the measurements as the fraction of overlap volume between the surveys to the total volume of the two surveys combined. However, this approach obviously cannot take into account any possible negative correlations between the data as the effect of the correlations can be due to more than the overlapping survey volumes. Thus, this approach can lead to a potentially biased covariance matrix. This issue will be resolved in the near future when upcoming surveys like Euclid and LSST will provide consistent growth-rate measurements in both the low and high redshift regimes.

\begin{figure*}[!t]
\centering
\includegraphics[width = 0.98\textwidth]{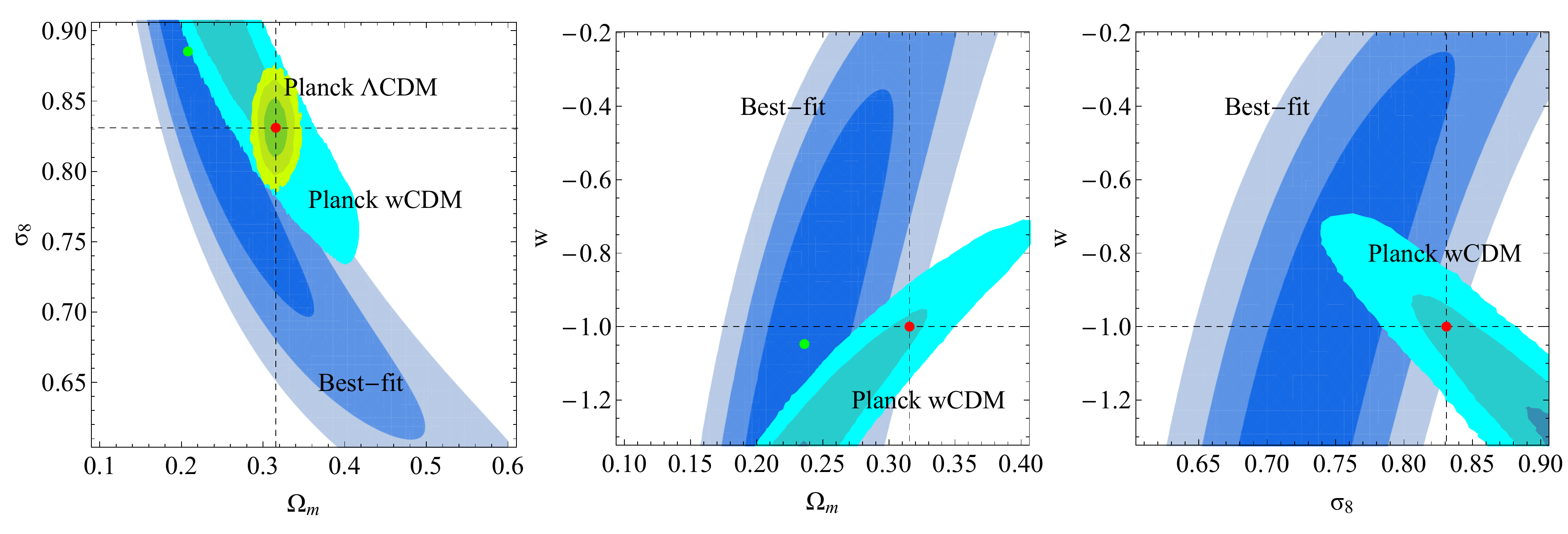}
\caption{The $68.3\%,~95.4\%$, and $99.7\%$ confidence contours in the $(w,\sigma_8,\Omega_{0m})$ parameter space. The red point corresponds to the \plcdm best-fit cosmology, the blue contour corresponds to the best-fit of our data to the \lcdm model (left) and $w$CDM model (middle and right panel), while the light blue and light green contours correspond to the Planck15 $w$CDM and \lcdm contours respectively. As we can see, there is a $3\sigma$ tension between the \plcdm values and the growth-rate data best-fit. }
\label{fig:contours}
\end{figure*}

Using the corrected $\chi^2$ and our ``Gold-2017" compilation given by Table \ref{tab:fs8-data-gold}, we now proceed to extract the best-fit cosmological parameters and discuss the results. First, we assume GR with a constant $w$ model and a flat Universe. Then, the Hubble parameter is given by
\ba
E(a)^2&\equiv & H(a)^2/H_0^2 \nn \\
&=& \Omega_{0m}a^{-3}+\left(1-\Omega_{0m}\right)a^{-3(1+w)},
\label{eq:Hubblew}
\ea
where we have ignored the radiation as at late times it has a negligible impact. This case is rather simple, so in order to speed up the code, it is convenient to use the analytical expression for the growing mode of the growth factor given by Eq.~(\ref{Da1}) and the analytical expression for the luminosity distance, which follows after a quick calculation using the definition, given by
{\small{\ba
\frac{H_0}{c}d_L(a)&=& \frac{2}{a \sqrt{\Omega_{0m}}}\, _2F_1\left(\frac{1}{2},-\frac{1}{6 w};1-\frac{1}{6 w};1-\frac{1}{\Omega_{m}(1)}\right) \nn \\
&-&\frac{2}{\sqrt{a} \sqrt{\Omega_{0m}}} \, _2F_1\left(\frac{1}{2},-\frac{1}{6 w};1-\frac{1}{6 w};1-\frac{1}{\Omega_{m}(a)}\right), \nn \\
\ea}}
where $\Omega_{m}(a)=\frac{\Omega_{0m}a^{-3}}{E(a)^2}$, and then the angular diameter distance is given by $d_A(z)=\frac{d_L(z)}{(1+z)^2}$ as usual. In the more complicated cases discussed in the next sections, \eg modified gravity models, we will perform the corresponding calculations numerically.

After fitting the data we obtain the $68.3\%,~95.4\% $ and $99.7\%$ confidence contours in the $(w,\sigma_8,\Omega_{0m})$ parameter space, shown in Fig.~\ref{fig:contours}. As it can be seen, the current growth rate data are at a $\sim3\sigma$ tension with the \plcdm best-fit cosmology, indicated with the red dot. For completeness we also overlap the corresponding Planck15/$w$CDM contours even though our goal here is to identify the tension level with \plcdm. We will attempt to alleviate this tension in the next section, by considering modified gravity models, as the extra degrees of freedom provided by the theories may allow a Newton constant of the form $G_{\textrm{eff}}(a,k)$ to account for the tension.

Remarkably, we find that compared to previous studies, \eg \cite{Basilakos:2016nyg} or even the Planck 2015 data release \cite{Ade:2015xua-planck-2015-params}, all of which use outdated growth data, with our new `Gold-2017' compilation we identify a $3\sigma$ tension. Given the \plcdm background and the fact that we have corrected for the diverse fiducial cosmologies used, this tension could potentially be explained either by assuming that the growth rates $f\sigma_8$ suffer from a yet unaccounted for systematic or by new physics perhaps affecting either the background $H(z)$ or inducing an evolution of Newton's constant due to modifications of GR.

In this paper, we will focus on the latter possibility and explore the various possibilities afforded by the rich phenomenology of modified gravity. As mentioned above, in these theories, Newton's constant can be time and scale dependent, \ie $G_\text{eff}(a,k)$, thus affecting the evolution of the growth factor via the last term in Eq.~(\ref{eq:ode}). We discuss these models in the what follows.

\section{Releasing the Tension using Modified Gravity}
\label{sec:Section 3}

In this section, we discuss physically motivated parametrizations of Newton's constant $G_{\textrm{eff}}(a,k)$, paying special attention to scale independent parametrizations motivated by modified gravity  theories on subhorizon scales. We first consider one of the minimal extensions of GR, the well-known $f(R)$ theories, where it maybe shown that under the subhorizon/quasi-static approximation \cite{Tsujikawa:2007gd}
\ba
\Geff/G_{\textrm{N}}&=&\frac{1}{F}\frac{1+4\frac{k^2}{a^2}m}{1+3\frac{k^2}{a^2}m} \label{gefffr1}\\ m&\equiv& \frac{F_{,R}}{F}\\F&\equiv&f_{,R}=\frac{\partial f}{\partial R}\label{gefffr2}
\ea
which reduces to GR only when $f(R)=R-2\Lambda$, \ie the \lcdm model, while a more accurate approximation was found in Ref.\cite{delaCruzDombriz:2008cp}.

One of the simplest extensions of GR and the \lcdm model with this formalism is the popular $f(R)$ model of Hu and Sawicki \cite{Hu:2007nk}. However, the original form of this model is unnecessarily complicated and has several degenerate parameters, so here we prefer the implementation of the $b$ parameter as in Ref.~\cite{Basilakos:2013nfa}. This has several useful advantages: first, the deviation of this model from $\Lambda$CDM is more transparent and second, by performing a Taylor expansion around $b=0$ we can obtain analytical approximations for $H(z)$ which are accurate to better than $0.1\%$ for $b\lesssim1$ and better than $10^{-5}\%$ for $b\lesssim0.1$. The Lagrangian for the Hu and Sawicki model, as written equivalently in Ref.~\cite{Basilakos:2013nfa}, is
\be
\label{Hu1}
f(R)= R- \frac{2\Lambda }{1+\left(\frac{b \Lambda }{R}\right)^n}
\ee
where $n$ is a constant of the model, usually chosen as $n=1$ without loss of generality as it only adjusts the steepness of the deviation from the \lcdm model.

As mentioned, we can also obtain a very accurate Taylor expansion of the solution to the equations of motion around $b=0$, i.e. the \lcdm model, as
\be
H^2(a)=H_{\Lambda}^2(a)+\sum_{i=1}^M b^i \delta H_i^2(a), \label{expansion1}
\ee
where
\be
\frac{H_{\Lambda}^2(a)}{H_0^2}=\Omega_{0m}a^{-3}+\Omega_{r0} a^{-4}+(1-\Omega_{0m}-\Omega_{r0})\label{LCDM1}
\ee and $M$ is the number of terms we keep before truncating the series. However, we have found that keeping only the two first non-zero terms is more than enough to have better than $0.1\%$ accuracy with the numerical solution. The functions $\delta H_i^2(a)$ are just algebraic expressions and can be easily determined from the equations of motion (see Ref.~\cite{Basilakos:2013nfa}). Finally, we also follow Ref.~\cite{Basilakos:2013nfa} and set $k=0.1h {\rm \textrm{Mpc}}^{-1}\simeq 300 H_0$, which is necessary as now the Newton's constant depends on the scale $k$ as well.

One can generalize the above model to an action that includes a scalar field with arbitrary kinetic term non-minimally coupled to gravity\footnote{Of course one can also consider other types of theories like models with Galileons, or with torsion of the type $f(T)$, non-minimal couplings and so on that have a similar effect. For this paper we limit ourselves to $f(R)$ and scalar-tensor theories in order to keep the problem tractable.}. Such a model has the following action \cite{Tsujikawa:2007gd}
\be
S=\int d^4x \sqrt{-g}\left(\frac12f(R,\phi,X)+\mathcal{L}_m\right),\label{eq:mogaction}
\ee
where $X=-g^{\mu\nu}\partial_\mu \phi \partial_\nu \phi$ is the kinetic term of the scalar field. In this case, Newton's constant is given by \cite{Tsujikawa:2007gd}:
\be
\Geff(a,k)/G_{\textrm{N}}=\frac{1}{F}\frac{f_{,X}+4\left(f_{,X} \frac{k^2}{a^2}\frac{F_{,R}}{F}+\frac{F_{,\phi}^2}{F}\right)}{f_{,X}+3\left(f_{,X} \frac{k^2}{a^2}\frac{F_{,R}}{F}+\frac{F_{,\phi}^2}{F}\right)},\label{geffmog}
\ee
where $F=F(R,\phi,X)=\partial_R f(R,\phi,X)$ and $F_{,\phi}=\partial_\phi F(R,\phi,X)$. This class of theories encompasses both the $f(R)$ models and the so called scalar-tensor (ScT) ones, given by the Lagrangian:
\be
\mathcal{L}^{\textrm{ScT}}=\frac{F(\phi)}{2}R+X-U(\phi) \label{eq:SCTEN}
\ee
and in this case Newton's constant reduces to
\be
\Geff(a,k)/G_{\textrm{N}}=\frac{1}{F(\phi)}\frac{F(\phi)+2F_{,\phi}^2}
{F(\phi)+\frac32F_{,\phi}^2}.\label{geffSCT}
\ee
It may be shown that on subhorizon scales both (\ref{geffmog}) and (\ref{geffSCT}) are well approximated by scale-independent functions. Thus, in what follows, we ignore the scale dependence of $G_{\textrm{eff}}$.

The effective Newton constant $G_{\textrm{eff}}$ can be related to the FRW metric perturbations and in particular to the Newtonian potentials $\Phi$ and $\Psi$ as in Ref.~\cite{Mueller:2016kpu}, \ie
\ba
ds^2&=&a^2\left[-(1+2\Psi)d\tau^2+(1-2\Phi)d\vec{x}^2\right], ~~~~~\\
\nabla^2 \Psi &=& 4 \pi G_{\textrm{N}} \rho \delta  \times G_M, \\
\nabla^2 (\Phi+\Psi) &=& 8 \pi G_{\textrm{N}} \rho \delta  \times G_L,
\ea
where $\delta$ is the growth factor and $G_L$ and $G_M$ are dimensionless parameters, which are equal to 1 in GR, but otherwise can be parametrized as functions of the scale factor in a variety of ways \cite{Mueller:2016kpu}. In this case, $G_M=G_{\textrm{eff}}/G_{\textrm{N}}$ alters the growth of matter, while $G_L$ alters the lensing of light via the lensing potential $\Phi+\Psi$.
Deviations from GR are also described through the gravitational slip defined as
\be
\gamma_{\textrm{slip}}=\frac{\Phi}{\Psi}
\ee
and through the anisotropic stress, that is inherent to most modified gravity theories and is defined as
\be
\eta=\frac{\Psi-\Phi}{\Phi}. \label{eq:eta}
\ee
Clearly, the gravitational slip and the anisotropic stress are related via $\eta=\frac{1}{\gamma_{\textrm{slip}}}-1$, and in GR, we have that $\gamma_{slip}=1$ and $\eta=0$. In Ref.~\cite{Tsujikawa:2007gd}, it was shown that in scalar-tensor theories the anisotropic stress is given by
\be
\eta=\frac{F^2_{,\phi}}{F(\phi)+F^2_{,\phi}}\label{eq:eta1}
\ee
and Eq.~\eqref{eq:eta} implies that the quantities $G_L$ and $G_M$ are related via
\be
G_L=\frac12 \frac{\eta+2}{\eta+1} G_M. \label{eq:GMGL}
\ee

In order to have agreement with the Solar System tests viable models must satisfy $F_{,\phi}\simeq 0$ at $z\simeq 0$, which from Eqs.~\eqref{eq:eta} and \eqref{eq:eta1} implies that $\eta \simeq 0$. Similarly, from \eqref{eq:GMGL}, we infer that at $z\simeq 0$, $G_L\simeq G_M \simeq 1$ and $\gamma_{\textrm{slip}}\simeq 1$.

Any of the above quantities $G_M$, $G_L$, $\gamma_{\textrm{slip}}$, or $\eta$ can be used to construct a null test for GR. Alternative approaches like the growth index\cite{Linder:2005in-growth-rate-history,DiPorto:2007ovd,Nesseris:2007pa-growth-analysis,Polarski:2007rr,Gannouji:2008jr,Gannouji:2008wt,Gong:2008fh,Kwan:2011hr} can also be used for parametrizing deviations from GR. However, they are not as efficient in distinguishing the effects of the background $H(z)$ from the effects of modified gravity since they do not enter explicitly in the dynamical growth equations.

In the present analysis we focus on $G_M=G_{\textrm{eff}}/G_{\textrm{N}}$ to parametrize deviations from GR since this is the only quantity that enters directly in the dynamical equation that determines the growth of density perturbations (eq (\ref{eq:ode})).  We thus use the parametrization
\ba
\frac{G_{\textrm{eff}}(a,n)}{G_{\textrm{N}}} &=& 1+g_a(1-a)^n - g_a(1-a)^{2n} \nn \\
&=&1+g_a\left(\frac{z}{1+z}\right)^n - g_a\left(\frac{z}{1+z}\right)^{2n}. \label{geffansatz}
\ea
Clearly, this parametrization mimics the large $k$ limit of the above models, \ie scales small compared to the horizon, which is a reasonable approximation even for large surveys.
In addition, the parametrization (\ref{geffansatz}) may be viewed as an extended Taylor expansion around $a=1$ for a fixed number of two parameters. The second term describes $G_\text{eff}$ for low and intermediate values of $z$, while the third term describes $G_\text{eff}$ for larger values of $z$. A similar parametrization concerning the dark energy equation of state was introduced in Ref.~\cite{Pantazis:2016nky}. The parametrization (\ref{geffansatz}) is only viable for $n\ge2$ due to the Solar System tests that demand that the first time derivative of $G_\text{eff}$ should be zero.

However, at this point it is important to mention that in the context of our analysis, we have assumed that the value of the effective Newton constant $G_\text{eff}$ is independent of the presence of matter density. Thus, the value of $G_\text{eff}$ on subhorizon scales is assumed to be scale and environment independent. We anticipate this assumption to be a good approximation in modified gravity models where in the physical frame there is no direct coupling of the scalar degree of freedom to matter density. In chameleon scalar-tensor field models, this assumption is not applicable, and therefore in such models, Solar System constraints are much less stringent, and our  parametrization with $n=1$ could be physically relevant. Therefore, in what follows, we will consider all values of $n$ with $n\ge0$.

Furthermore, this parametrization is motivated by considering that any viable modified gravity model must satisfy the following conditions:
\begin{itemize}
  \item $G_{\textrm{eff}}>0$ in order for the gravitons to carry positive energy.
  \item $G_{\textrm{eff}}/G_{\textrm{N}}=1.09\pm 0.2$ to be in agreement with the Big Bang Nucleosynthesis (BBN).
  \item Today, we should have $G_{\textrm{eff}}(a=1)/G_{\textrm{N}}=1$ due to our choice for the normalization of $F$.
\end{itemize}

As can be seen then, for $g_a>-4$, our parametrization of Eq.~(\ref{geffansatz}) satisfies all of the aforementioned requirements \footnote{ Note that for $g_a<0$, the parametrization of Eq.~(\ref{geffansatz}) has a minimum at $a_{G_\text{eff,min}}=1-2^{-1/n}$ with a value of $G_{\textrm{eff}}/G_{\textrm{N}}=1+\frac{g_a}{4}$, hence in order to have $G_{\textrm{eff}}>0$ we need $g_a>-4$ and as we will see later on, the best-fit for various $n$ satisfies that.}. One could also demand that at early times we have $G_{\textrm{eff}}'(a=0)/G_{\textrm{N}}=1$, \ie we have GR, but that would require yet another term in Eq.~(\ref{geffansatz}), so that the coefficient can adjust the first derivative, but since the BBN constraint is not so stringent, we prefer to allow for some extra freedom in our model.

Another criterion we should take into account is the self-consistency of the modified gravity model, as not all parametrizations can be reproduced by a given model. For example, in Ref.~\cite{Nesseris:2006jc}, it was found that a scalar-tensor model cannot reproduce a given combination of $G_{\textrm{eff}}(z)$ with $H(z)$. This was manifested by a negative kinetic term for the scalar field $\Phi$; see Fig.~5 in Ref. \cite{Nesseris:2006jc}.\footnote{For a similar reason GR-quintessence does not allow crossing of the phantom divide line $w=-1$ as this crossing would require a change of sign of the scalar field kinetic term.} Thus, we are lead to the following question: what are the allowed $H(z)-G(z)$ regions for a given modified gravity model? For scalar-tensor theories and $H(z)$ corresponding to \lcdm, this question is addressed in the next section.

\begin{figure}[!t]
\centering
\includegraphics[width = 0.48\textwidth]{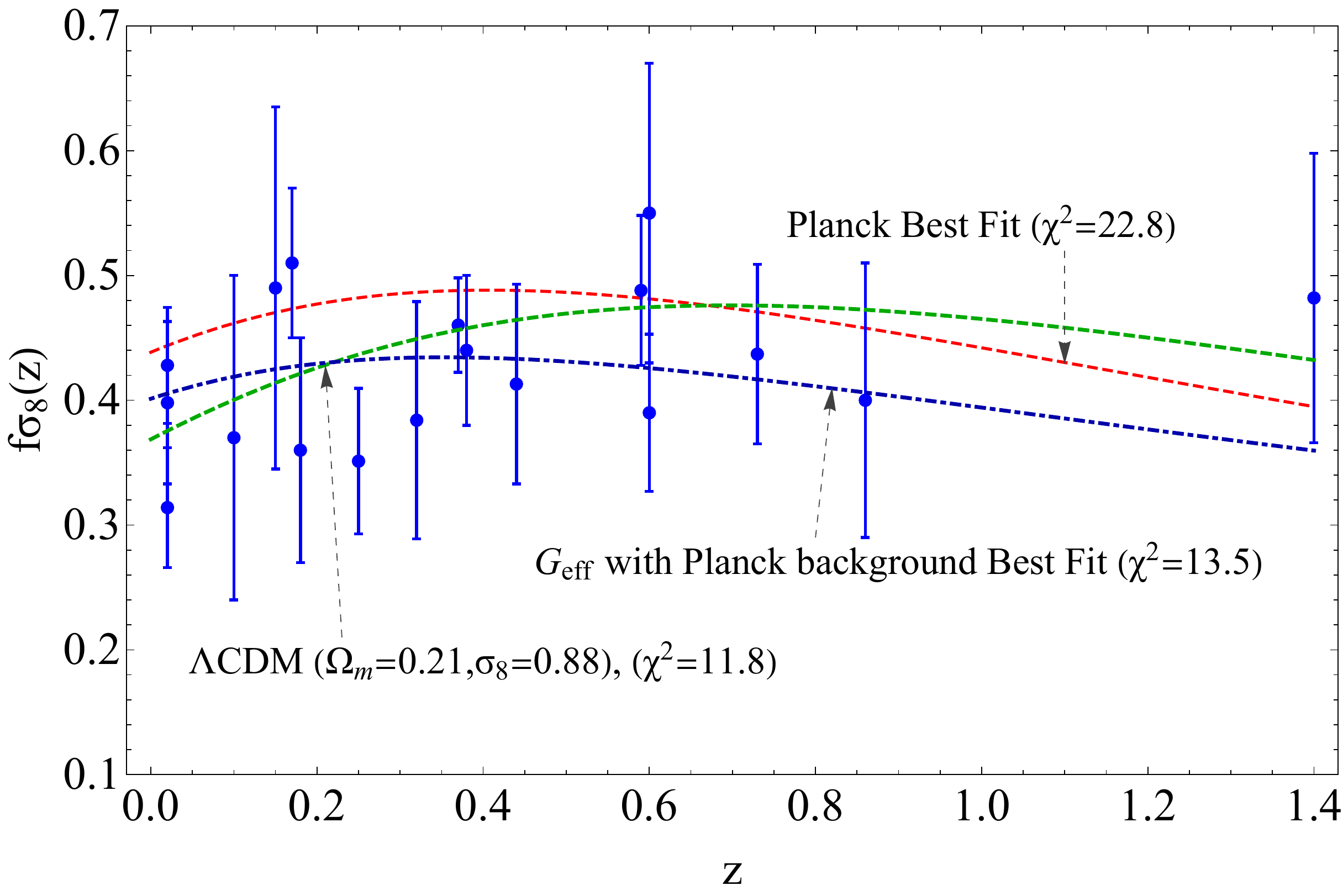}
\caption{Plot of $f\sigma_8(z)$ for the `Gold-2017' growth rate dataset. The green dashed line and the red dashed one correspond to the best fits of \lcdm and \plcdm models respectively, while the blue dot-dashed one corresponds to the best fit of $G_\text{eff}$ parametrization for $g_a = -1.16, n=2$ with the Planck15 background.}
\label{fig:fs8z}
\end{figure}

\begin{figure}[!t]
\centering
\includegraphics[width = 0.48\textwidth]{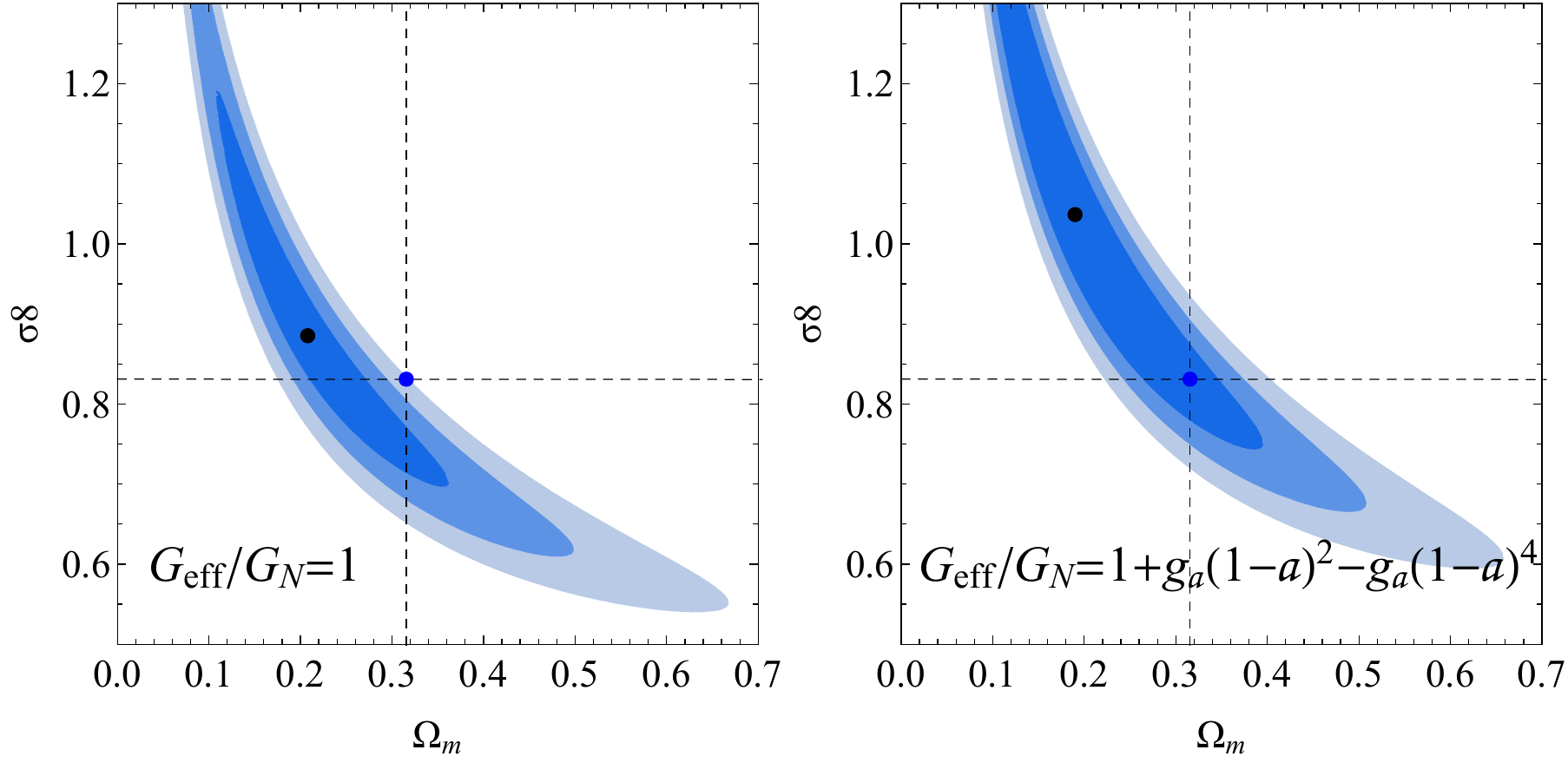}
\caption{The $68.3\%,~95.4\% $ and $99.7\%$ confidence contours in the $g_a-n$ plane for $n=0$ (left) and $n=2$ (right). In both cases, the point at the intersection of the dashed lines corresponds to the best-fit \plcdm  model.}
\label{fig:gan}
\end{figure}

\begin{figure}[!t]
\centering
\includegraphics[width = 0.48\textwidth]{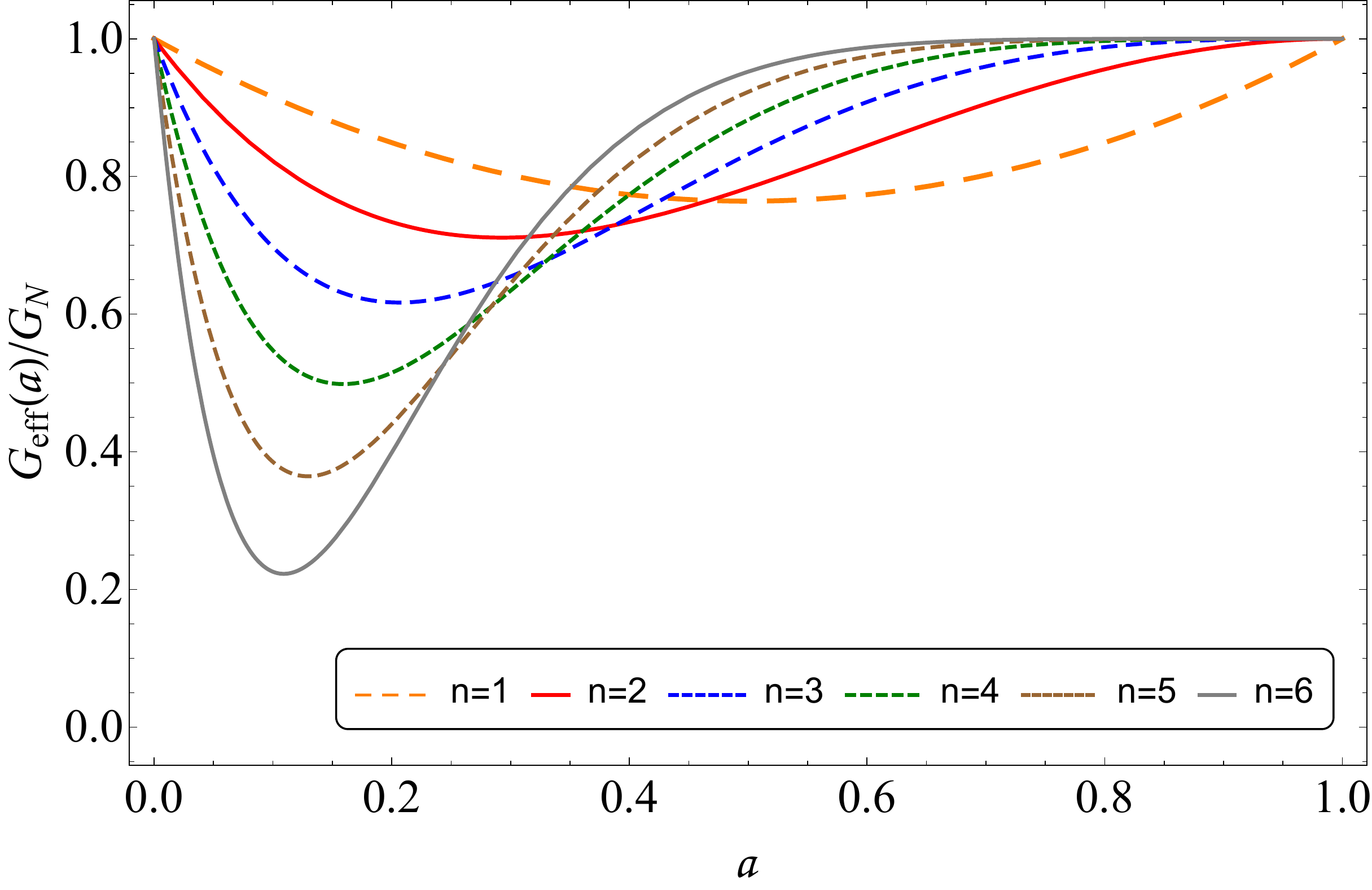}
\caption{$G_\text{eff}(a)/G_\text{N}$ in the range $n \in [1,6]$ from the upper to lower curve respectively and for $g_a$ corresponding to the best-fit value.}
\label{fig:geffa}
\end{figure}

In the rest of this section we fit the parametrization (\ref{geffansatz}) to our `Gold-2017' assuming a \plcdm background. The $f\sigma_8$ datapoints of the `Gold-2017' dataset along with some model fits are shown in Fig.~\ref{fig:fs8z}. The green dashed line corresponds to the best fit of the \lcdm model  ($\chi^2=11.8$, \omom $=0.21, \sigma_8 =0.88$), the red dashed one corresponds to the \plcdm model parameter values ($\chi^2=22.8$) and the blue dot-dashed one corresponds to the best fit of $G_\text{eff}$ parametrization for $(g_a = -1.16, n=2)$ with the Planck15 background ($\chi^2=13.5$). In particular, in Fig.~\ref{fig:gan} we show the $68.3\%,~95.4\% $ and $99.7\%$ confidence contours in the $g_a-n$ plane for $n=0$ and $n=2$, while in Fig.~\ref{fig:geffa} we show $G_{\textrm{eff}}(a)$ for various values of $n$ ($n \in [2,6]$ from the upper to the lower curve respectively) and for $g_a$ corresponding to the best-fit value. We note that for $a \ll 1$ the constraint from BBN is satisfied since $G_\text{eff}(a \ll 1)/G_{\textrm{N}}=1$.

Specifically, we explore systematically the parametric space for $n=1,2,\dots, 6$, while the best fit values of $g_a$ with error bars are shown explicitly in Table \ref{tab:geff_ga}. It is important to mention at this point that these best fit values refer to the first minimum of each $\chi^2$, which is not always the global one. We discuss the issue of the multiple minima in some detail in Appendix~\ref{sec:Appendix_B}. As can be seen from Fig.~\ref{fig:gan}, this parametrization is capable of alleviating the tension found between the growth rate data and the \plcdm best fit, reducing it from $\sim3\sigma$ to less than $1\sigma$, thus offering potential hints for new physics.

As mentioned above, the consistency of any modified gravity model with the Solar System tests is paramount as they place stringent constraints on the evolution of $G_\text{eff}$. Hence, viable models like the Hu and Sawicki model \cite{Hu:2007nk}  that evade them  are effectively small perturbations around the \lcdm (see \eg  Eq.~(\ref{Hu1})). From a phenomenological point of view it is also interesting to consider direct parametrizations of $G_{\textrm{eff}}$ like the one of Eq.~(\ref{geffansatz}). Such a consideration leads to the following question: are the best forms of $G_{\textrm{eff}}$ able to lead to a reconstruction of self- consistent scalar-tensor quintessence with the \plcdm background? We will address this question in the next section.\\

\begin{table}[!t]
\begin{centering}
\caption{The best fit values of $g_a$ with errors bars for $n=1,2,\dots, 6$. As we describe in Appendix~\ref{sec:Appendix_B}, this parametrization has several distinct minima, but here we show only the global one when both $g_a$ and $n$ are free (first row) and then for integer values of $n = 1, 2, \dots, 6$, the minima corresponding to the lowest $g_a$ which are also the global ones for low values of $n$.
\label{tab:geff_ga}}
\begin{tabular}{cc}
\hline
$n$ & $g_a$ \\
\hline  \hline
$0.343$ & $-1.200 \pm 1.025$ \\
$1$ & $-0.944 \pm 0.253$ \\
$2$ & $-1.156 \pm 0.341$ \\
$3$ & $-1.534 \pm 0.453$ \\
$4$ & $-2.006 \pm 0.538$ \\
$5$ & $-2.542 \pm 0.689$ \\
$6$ & $-3.110 \pm 0.771$ \\
\hline
\end{tabular}
\end{centering}
\end{table}

\section{Reconstruction of Scalar-Tensor Quintessence}
\label{sec:Section 4}

The line element for the FLRW metric corresponding to a flat universe is given by
\be
ds^2 = -dt^2 + a^2(t) \left[dr^2 + r^2 (d\theta^2 + \sin^2\theta \ d\phi^2) \right].
\ee
Using this metric in the action (\ref{action}) and assuming a homogeneous scalar field and a perfect fluid background we find the dynamical equations of the system as
\begin{eqnarray}
3F H^2 &=&  \rho +{1\over 2} \dot\phi^2 - 3 H \dot F + U \label{fe1}\\
-2F \dot H  &=& (\rho+p) + \dot \phi^2 +\ddot F - H \dot F \label{fe2} \label{dmt}
\end{eqnarray}

We eliminate the kinetic term $\dot{\phi}^2$ in Eq.~\eqref{fe2}, and we set the squared rescaled Hubble parameter as
\be
q(z) \equiv E^2(z) = \frac{H^2(z)}{H_0^2},
\ee
while a new rescaling to potential is applied, \ie $U \rightarrow U \cdot H_0^2$. We thus obtain the dynamical equations in terms of the redshift $z$ as
\begin{widetext}
\ba F^{\prime\prime}(z) &+& \left[\frac{q^\prime(z)}{2q(z)}-\frac{4}{1+z}\right] F^{\prime}(z) + \left[\frac{6}{(1+z)^2} - \frac{2}{(1+z)}\frac{q^\prime(z)}{2q(z)}\right] F = \frac{2U(z)}{(1+z)^2 q(z)} + 3 \frac{1+z}{q(z)} \Omega_{0m} \ \label{fe1a} \\ \phi^{\prime}(z)^2 &=& -\frac{6 F^\prime(z)}{1+z} + \frac{6 F(z)}{(1+z)^2} -\frac{2 U(z)}{(1+z)^2 q(z)} - 6 \frac{1+z}{q(z)} \Omega_{0m} \label{fe2a} \ea
\end{widetext}
where the differentiation with respect to the redshift $z$ is denoted by the prime and we have assumed a matter perfect fluid with $p=0, \Omega_{0m} = 3 \rho_{0m}/ H_0^2$. In addition, Eqs. \eqref{fe1a} and \eqref{fe2a} satisfy the initial conditions $\phi(0)=0, \ F(0)=1$, and $F^\prime(0)=0$ for consistency with Solar System tests $(dF/d\phi \sim dF/dz\simeq0$ \cite{Will:2005va,Scharre:2001hn,Poisson:1995ef}).

In scalar-tensor theories, the effective Newton constant with respect to $z$ is of the form (see Ref. \cite{Nesseris:2006jc})
\be
G_{\text{eff}}(z) = \frac{1}{F} \frac{2F + 4 \left(\frac{dF}{d\Phi} \right)^2}{2F + 3 \left(\frac{dF}{d\Phi} \right)^2} \ G_\text{N} \simeq \frac{G_\text{N}}{F},
\label{Geff}
\ee
where $G_\text{N}$ is the well known Newton constant in GR.

Equations \eqref{fe1a} and \eqref{fe2a} form the system of equations for $\{U(z), \phi^\prime(z) \}$ that can be used for the reconstruction of the theory (derivation of functions $U(\phi)$, $F(\phi)$), assuming that the functions $F(z)$ (or $G_{\textrm{eff}}(z)$) and $H(z)$ are observationally obtained  (\cite{Chevallier:2000qy-cpl-param1,Sahni:2014ooa-bao-anomaly1,Nesseris:2006er}). The function $H(z)$ is well approximated by the \plcdm fit with parameters shown in Table \ref{tab:planck}. The function $G_{\textrm{eff}}(z)$ may be obtained using the growth data of Table \ref{tab:fs8-data-gold} in the context of the parametrization \eqref{geffansatz} that satisfies the three basic conditions discussed in the previous section (Solar System tests, nucleosynthesis constraints and proper normalization at the present time).

Even after the observational determination of $G_{\textrm{eff}}(z)$ and $H(z)$ the self-consistent reconstruction of a modified theory is not always possible. For example, in the case of a scalar-tensor theory the sign of $\phi^{\prime}(z)^2$ obtained from Eqs.~\eqref{fe1a} and \eqref{fe2a} may turn out to be negative leading to a complex predicted value of the scalar field. This violates that assumption of a real scalar field on which the theory is based and leads to inconsistencies that may be difficult to overcome.

\begin{figure}[!t]
\centering
\includegraphics[width = 0.48\textwidth]{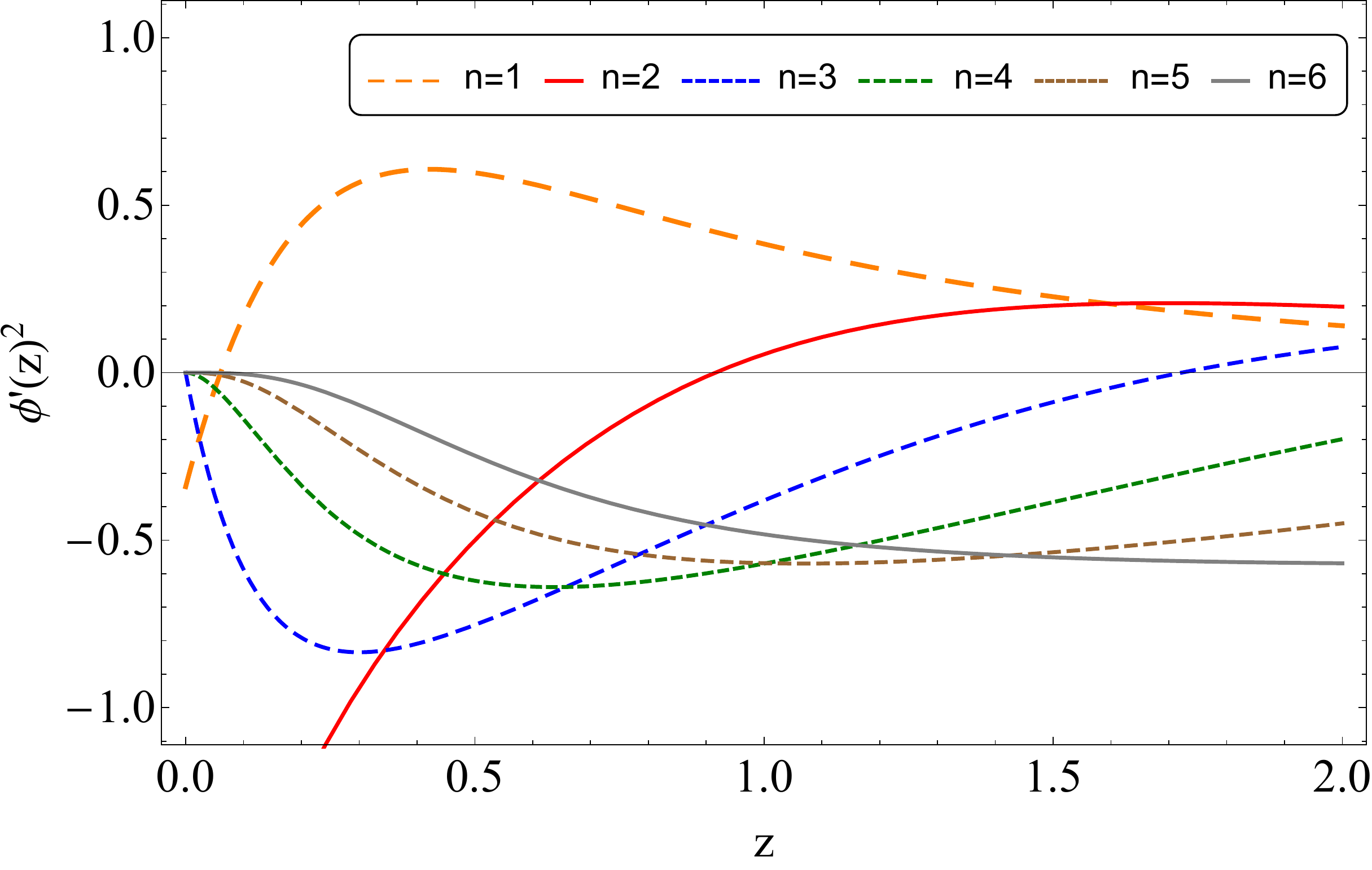}
\caption{The evolution with redshift of the kinetic term $\phi^{\prime}(z)^2$ for various values of $n$ in the range $z \in [0,2]$. Each case gives an imaginary scalar field which is not acceptable leading to scalar-tensor theory inconsistencies. The line corresponding to $n=1$ is only applicable in a chameleon mechanism and thus is ruled out due to Solar System tests in the present analysis.}
\label{fig:phi2prime}
\end{figure}

As shown in Fig. \ref{fig:geffa} and in Table \ref{tab:geff_ga} it is clear that the growth data indicate that the gravitational strength may be a decreasing function of the redshift in the redshift range [0,0.4] compared to its present value. The question that we want to address is the following: can this weakening effect of gravity be due to an underlying scalar-tensor theory?  If the answer is positive, then the sign of the reconstructed $\phi^{\prime}(z)^2$ should be positive so that the scalar field of the theory is real. We will show that a $G_{\textrm{eff}}$ that is decreasing with redshift at low $z$ is not consistent with positive $\phi^{\prime}(z)^2$ and therefore this behavior cannot be due to an underlying scalar-tensor theory. This is shown numerically in Fig. \ref{fig:phi2prime} where we show the reconstructed form of $\phi^{\prime}(z)^2$ under the assumption of the best fit forms of $G_{\textrm{eff}}$ $(n=1,2,3,4,5,6)$ shown in Fig. \ref{fig:geffa} and the \plcdm background $H(z)$ obtained with the parameters of Table \ref{tab:planck}. Clearly, for all values of $n$ considered, $\phi^{\prime}(z)^2$ is negative for low $z$, leading to an unacceptable scalar-tensor theory.

This result may be generalized analytically as follows: using Eqs.~\eqref{fe1a} and \eqref{fe2a} and demanding that $\phi^{\prime 2} (z) \ge 0$, we obtain
\ba
F''(z)&+& F'(z) \left(\frac{q'(z)}{2q(z)}+\frac{2}{z+1}\right) -F(z)\frac{q'(z)}{(z+1) q(z)}\nn \\
&+&\frac{3 \Omega_m (z+1)}{q(z)}\leq0, \label{eq:difineq1}
\ea
which is a second-order differential inequality for $F(z)$.

A useful theorem for dealing with such inequalities is the Chaplygin theorem (see Ref.~\cite{Chaplygintheorem} and Appendix~\ref{sec:Appendix_A} for details). In order to bring the inequality (\ref{eq:difineq1}) to the form required by the theorem, we first set $F(z)=1-\delta f(z)$ and deduce the corresponding inequality for $\delta f(z)$. We then find
\ba
\delta f''(z)&+&\delta f'(z)\left(\frac{2}{1+z}+\frac{q'(z)}{2q(z)}\right)- \delta f(z)\frac{q'(z)}{(1+z) q(z)}\nn \\
&-&\frac{3 \Omega_m (1+z)}{q(z)}+\frac{q'(z)}{(1+z) q(z)}\geq 0. \label{eq:difineq2}
\ea
By applying the theorem, as described in Appendix~\ref{sec:Appendix_A}, we find that the inequality (\ref{eq:difineq1}) is satisfied for an \lcdm background only when $\delta f(z)\geq 0$ or $F(z)\leq1$ $(G_\textrm{eff}(z)/G_{\textrm{N}} \ge 1)$ for a range that includes all $z\geq0$, as we found by a numerical analysis.

To summarize, in order to satisfy the inequality (\ref{eq:difineq1}) along with the viability constraints (positive energy for the graviton \etc) and to be able to reconstruct the scalar-tensor Lagrangian in a \lcdm background, we need to have $0 < F(z)\leq 1$ $(G_\textrm{eff}(z)/G_{\textrm{N}} \ge 1)$. This result explains why the reconstruction as seen in Fig.~\ref{fig:phi2prime} does not work. Every one of these cases has a negative value for $\phi^{\prime}(z)^2$ at some $z$ and, as seen in Fig.~\ref{fig:geffa}, it also  has $F(z)>1$ $(G_\textrm{eff}(z)/G_{\textrm{N}} < 1)$ in some region. Several numerical tests we performed with several models seem to corroborate the result of this theorem. This issue has also been discussed in Ref.\cite{EspositoFarese:2000ij} even though no general rule was derived for the viability of the reconstruction. Therefore, we conclude that the only viable models of scalar-tensor theories that can be reconstructed in the context of a \lcdm background $H(z)$ are the ones where the non-minimal coupling function satisfies $0< F(z)\leq1$ for all $z\geq0$. In the context of the reconstruction analysis we have used the approximation that $G_{\textrm{eff}}\simeq \frac{1}{F}$, which we find is valid everywhere except when $\phi^{\prime}(z)^2$ changes sign.

\begin{figure*}[!t]
\centering
\includegraphics[width = 0.49\textwidth]{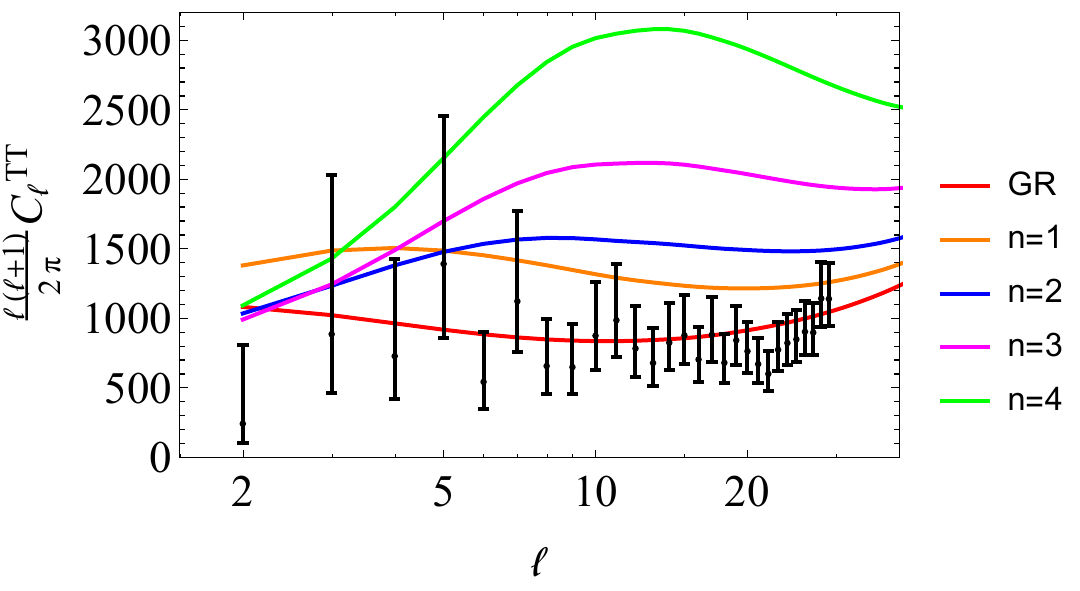}
\includegraphics[width = 0.49\textwidth]{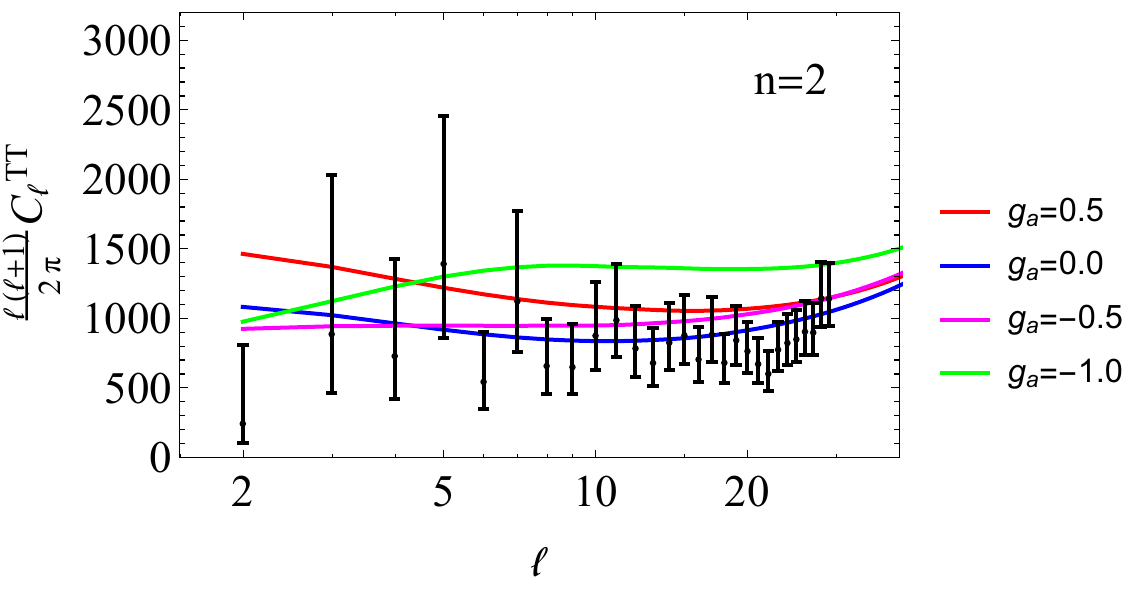}
\caption{The ISW effect for the $G_{\textrm{eff}}$ model used in our analysis for various values of $n$ evaluated at the minima for $g_a$ given in Table \ref{tab:geff_ga} (left) and for $n=2$ but for various values of $g_a$ as indicated by the label. We also show the Planck15 low-$\ell$ binned $C_\ell^{TT}$ data.}
\label{fig:ISW}
\end{figure*}

\section{Effects of $G_{\textrm{eff}}(z)$ on the CMB \label{sec:isw}}
In this section we investigate the effects of a redshift dependent $G_{\textrm{eff}}(z)$ on the CMB spectrum. We anticipate (and verify with MGCAMB below) that $G_{\textrm{eff}}(z)$ affects only the large angular CMB spectrum scales  (low-$\ell$) through the Integrated Sachs Wolfe (ISW) effect while smaller scales (the acoustic peaks) depend only on the background $H(z)$ through the angular diameter distance $d_A=\frac{c}{H_0}\frac{1}{1+z} \int_0^z \frac{1}{H(z')} dz'$. The ISW effect is significantly affected by the redshift dependence of $G_{\textrm{eff}}$ because it depends on the time evolution of the potential $\Phi(z)$ which in turn depends on $G_{\textrm{eff}}$ due to the Poisson equation $\frac{k^2}{a^2} \Phi(k,z)\propto \delta(z)\cdot G_{\textrm{eff}}(k,z)$, where $\delta=\frac{\delta \rho}{\rho}$ is the growth factor.

In Fig.~\ref{fig:ISW} we show a comparison of the theoretically predicted low-$\ell$ multipoles of the TT part of the CMB spectrum including the ISW effect for the best fit $G_{\textrm{eff}}$ models (Table \ref{tab:geff_ga}) (continuous lines left panel). The Planck15 low-$\ell$ binned $C_\ell^{TT}$ data are also shown. The theoretically predicted  spectra were obtained with a modified version of MGCAMB \cite{Hojjati:2011ix} with $ G_{\textrm{eff}}(z)/G_\textrm{N}$ given by \eqref{geffansatz}, anisotropic stress $\eta(z)=0$ and with the parameter values shown in Table \ref{tab:geff_ga} for $n=2,3,4$ and for $G_{\textrm{eff}}/G_{\textrm{N}} =1$  for GR. The right panel of Fig.~\ref{fig:ISW} shows the theoretically predicted CMB spectra for $n=2$ and various values of $g_a$.

Clearly, the higher the exponent $n$ of our parametrization for $G_{\textrm{eff}}$, the stronger the ISW effect and its deviation from the \lcdm model. Thus, the cases for $n=5,6$ are not included in Fig.~\ref{fig:ISW} as they are not consistent with the observed CMB power spectrum. As shown in Fig.~\ref{fig:geffa}, a higher $n$, means that gravitational strength varies more rapidly at low $z$ leading to the stronger ISW effect shown in Fig.~\ref{fig:ISW}.

\begin{figure}[!t]
\centering
\includegraphics[width = 0.49\textwidth]{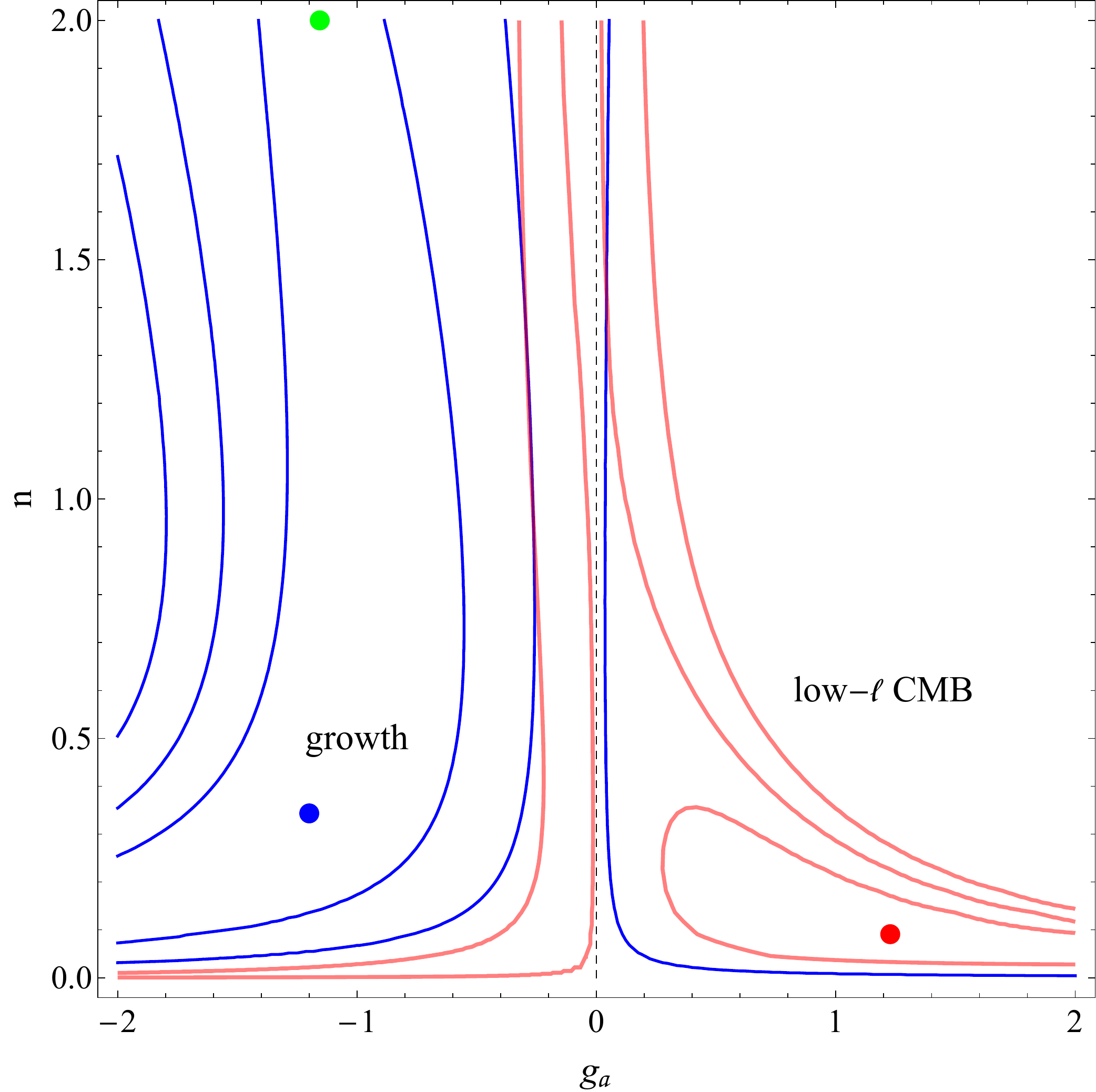}
\caption{The $1,2$ and $3\sigma$ contours for the $G_{\textrm{eff}}$ model in the $(g_a,n)$ parameter space based on the low-$\ell$ TT CMB data (red lines) and the growth rate data (blue lines). The black dashed line at $g_a=0$ but also the axis at $n=0$ correspond to GR and the \lcdm model, while the green, blue and red dots correspond to the best-fit for $n=2$, \ie $(g_a,n)=(-1.156, 2)$, the global minimum for $(g_a,n)=(-1.200, 0.343)$ and the minimum for the low-$\ell$ data, \ie $(g_a,n)=(1.227, 0.091)$, respectively. The blue and red contour regions are centered around the blue and red points respectively. As can be seen, there is a strong tension between the two datasets.}
\label{fig:contourgan}
\end{figure}

Also, we performed a simple $\chi^2$ analysis with the low-$\ell$ data, where we defined
\be
\chi_{low-\ell}^2=\sum^N_{i=1} \left(\frac{D_\ell^{Pl}-D_\ell^{th}}{\sigma_{D_\ell^{Pl}}}\right)^2 \label{lowl-like}
\ee
and $D_\ell=\frac{\ell (\ell+1)}{2\pi} C_\ell^{TT}$. In this case, we kept all other parameters except $g_a$ and $n$ fixed to their \plcdm values. We found that the \lcdm model ($n=0$ or $g_a=0$) has $\chi^2_{GR}=22.394$ and the rest of the models have $\chi^2_{n=2}=255.683$, $\chi^2_{n=3}=723.922$ and $\chi^2_{n=4}=2086.69$. Thus, these models are strongly disfavored with respect to \lcdm due to their rapid variation of $G_{\textrm{eff}}$ leading to strong effects on the ISW effect. In the case of fixed $n$, we find that $\chi^2_{g_a=0.5}=66.346$, $\chi^2_{g_a=0}=22.394$, $\chi^2_{g_a=-0.5}=42.755$ and $\chi^2_{g_a=-1}=186.969$, or in the case of $g_a=-0.5$, \ $\delta \chi^2 =20.361$, corresponding to a $4.1\sigma$ deviation. Thus the ISW effect provides significantly stronger constraints on $G_\text{eff} (z)$ than the growth data.

Our assumption that the probability distribution of the low $\ell$ TT CMB spectrum likelihood can be modeled as a Gaussian may not be accurate (see Ref.~\cite{Aghanim:2015xee}), and can introduce additional uncertainties which may reduce the tension level found in our analysis. Therefore, our $\chi^2$ analysis with the low-$\ell$ TT CMB data should be interpreted with extreme caution as it neglects the covariances of the data but also possible effects from the foregrounds which are clearly non-Gaussian. The effect of the non-Gaussianity will manifest itself as higher-order terms related to the skewness and the kurtosis, which has been shown to be relevant for the CMB (see eg  Ref.~\cite{Amendola:1994pq}).

\begin{figure*}[!t]
\centering
\includegraphics[width = 0.45\textwidth]{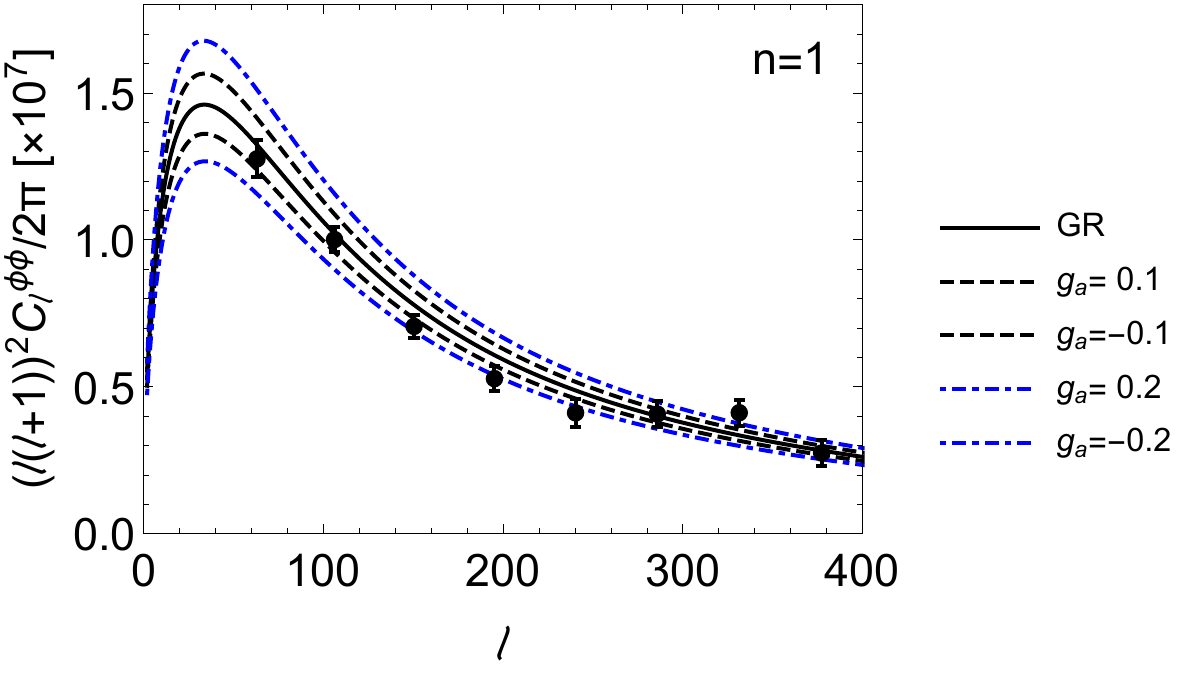}
\includegraphics[width = 0.45\textwidth]{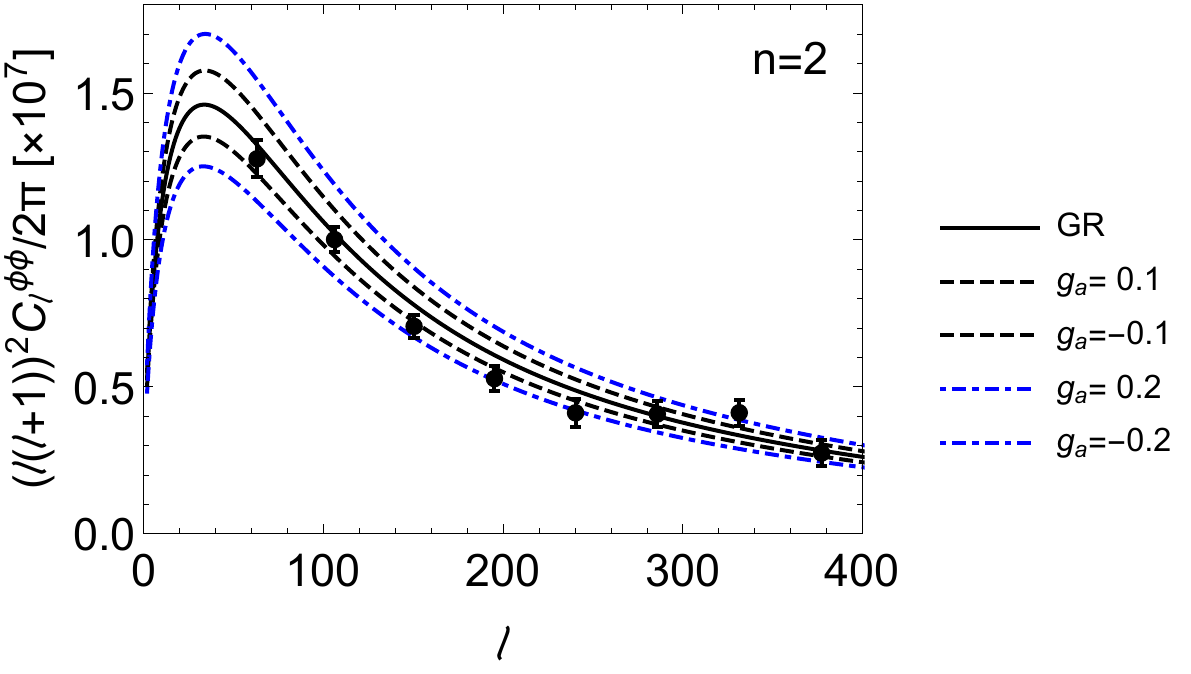}
\caption{The lensing potential for $\Lambda$CDM (black solid line) or the $G_{\textrm{eff}}$ model for $g_a=\pm0.1$ (black dashed line) and $g_a=\pm0.2$ (blue dot-dashed line). The data points are from Planck 2015 and were derived from the observed trispectrum\cite{Ade:2015zua}.}
\label{fig:cmblensing}
\end{figure*}

\begin{figure}[!t]
\centering
\includegraphics[width = 0.49\textwidth]{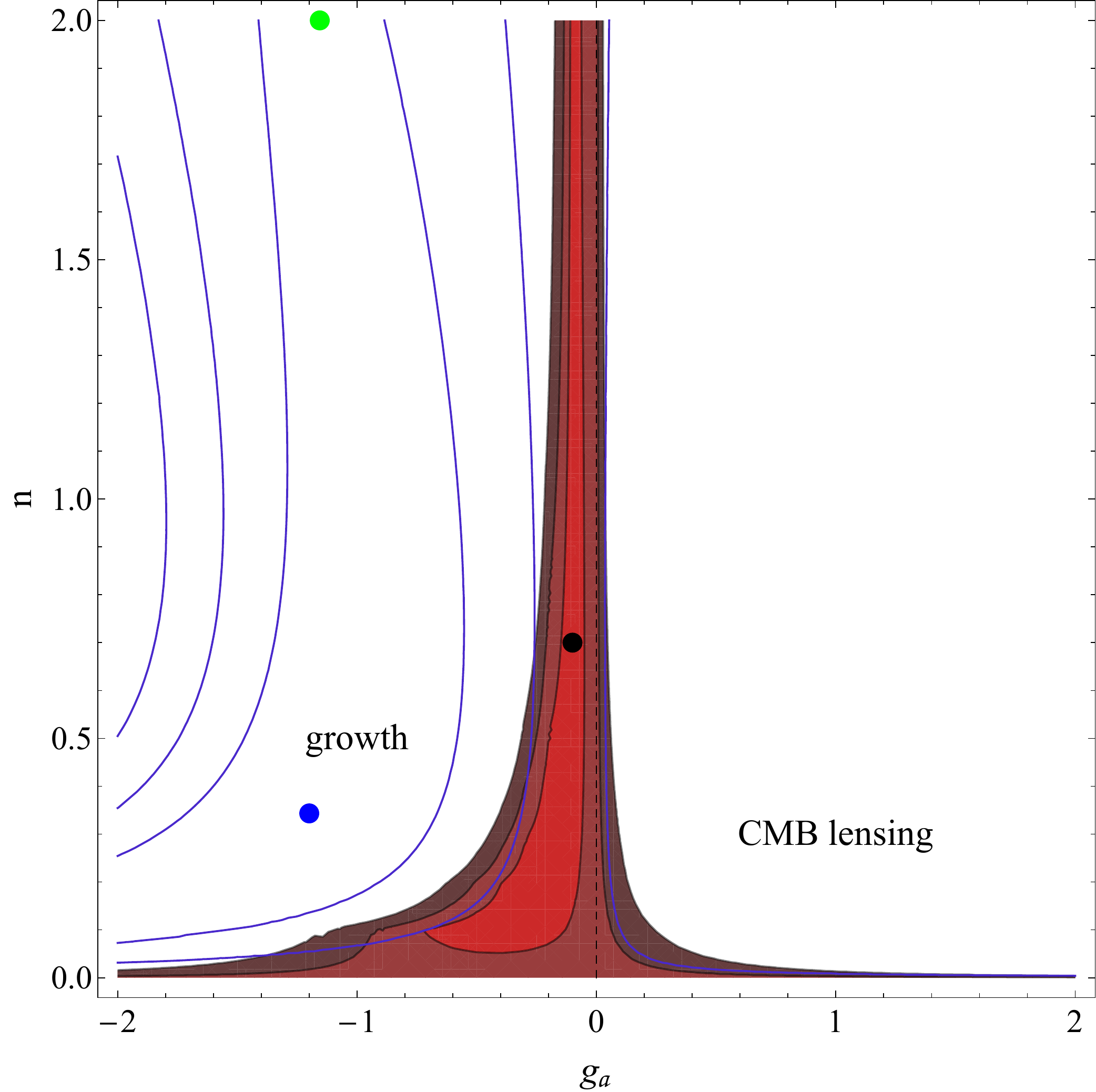}
\caption{The $1,2$ and $3\sigma$ contours for the $G_{\textrm{eff}}$ model in the $(g_a,n)$ parameter space based on the CMB lensing (trispectrum \cite{Ade:2015zua}) data (red contours) and the growth rate data (blue lines). The black dashed line at $g_a=0$ and also the axis at $n=0$ correspond to GR and the \lcdm model, while the green and blue points correspond to the best-fit for $n=2$, \ie $(g_a,n)=(-1.156, 2.000)$ and the global minimum for $(g_a,n)=(-1.200, 0.343)$ with the growth-rate data, while the black point to the CMB lensing best-fit for $(g_a,n)=(-0.200, 0.700)$. As can be seen there is a mild $2\sigma$ tension between the growth data contours (blue lines) and the CMB lensing contours (red lines).}
\label{fig:cmblensingcontours}
\end{figure}

In Fig.~\ref{fig:contourgan}, we also show the contours for the $G_{\textrm{eff}}$ model in the $(g_a,n)$ parameter space based on the low-$\ell$ TT CMB data (red lines) and the growth rate data (blue lines). The black dashed line at $g_a=0$ and the axis at $n=0$ correspond to GR and the \lcdm model since the last two terms in \eqref{geffansatz}in both cases cancel out. The green, blue and red dots correspond to the best-fit for $n=2$, \ie $(g_a,n)=(-1.156, 2)$, the global minimum for $(g_a,n)=(-1.200, 0.343)$ and the minimum for the low-$\ell$ data, \ie $(g_a,n)=(1.227, 0.091)$, respectively. Clearly, there is strong tension between the best fit growth data and the Planck low-$\ell$ power spectrum (ISW effect).

Another interesting probe to consider is the CMB lensing \cite{Ade:2015zua} which is sensitive to the impact of a modified growth rate. Clearly, modifications introduced by a time-dependent gravitational constant translate to significant changes in the CMB lensing. In this regard, in Fig.~\ref{fig:cmblensing}, we show the lensing potential for Planck15-$\Lambda$CDM (black solid line) along with the $G_{\textrm{eff}}$ model for $g_a=\pm0.1$ (black dashed line) and $g_a=\pm0.2$ (blue dot-dashed line). The data points are from Planck 2015 and were derived from the observed trispectrum\cite{Ade:2015zua}.

By fitting the modified lensing potentials for $\Lambda$CDM to the data we have also obtained new stronger constraints on the parameters of our parametrization. In Fig.~\ref{fig:cmblensingcontours} we show the $1,2$ and $3\sigma$ contours for the $G_{\textrm{eff}}$ model in the $(g_a,n)$ parameter space based on the CMB lensing (trispectrum \cite{Ade:2015zua}) data (red contours) and the growth rate data (blue lines). The black dashed line at $g_a=0$ and also the axis at $n=0$ correspond to GR and the \lcdm model, while the green and blue points correspond to the best-fit for $n=2$, \ie $(g_a,n)=(-1.156, 2.000)$ and the global minimum for $(g_a,n)=(-1.200, 0.343)$ with the growth-rate data, while the black point to the CMB lensing best-fit for $(g_a,n)=(-0.200, 0.700)$. As can be seen, there is a mild $2\sigma$ tension, and the allowed parameter space from the lensing potential data is significantly reduced and much more constraining than the ISW.

Finally, as mentioned above the $\chi^2$ analysis with the low-$\ell$ TT CMB data should be interpreted with extreme caution as it neglects the covariances and the non-Gaussianity of the data \cite{Aghanim:2015xee}. In addition all other parameters, such as $\Omega_m, H_0$ \etc, are fixed to their \plcdm values, so it would be worthwhile to do a full MCMC and explore the whole parameter space, which is left for future work. Thus, our analysis indicates that, even though the tension between the growth data and the \plcdm background in the context of GR is removed by allowing a redshift evolution of $G_{\textrm{eff}}(z)$, the required $G_{\textrm{eff}}(z)$ is not consistent with either scalar-tensor theories nor the low-$\ell$ CMB spectrum as determined by the ISW effect.

\section{Conclusions-Discussion \label{sec:Conclusion}}

We presented a collection of 34 growth rate data based on recent RSD measurements obtained from several surveys and studies over the last 10 years. In an effort to maximize robustness and independence of the data we selected 18 of the 34 growth rate data to construct a 'Gold-2017' growth rate dataset. Using this dataset we fit a $w$CDM cosmology and find that the best fit parameters $(w,\sigma_8,\Omega_{0m})$ are in $3\sigma$ tension with the corresponding parameters obtained with the Planck15 CMB data in the context of GR and \lcdm. In order to resolve this tension we consider a simple parametrization for $G_\text{eff}$ given by Eq.~(\ref{geffansatz}). We show that the tension in the parameters of the data gets now reduced to the $1\sigma$ level.

Despite this reduction of the tension between the growth data and the Planck indicated background, this best fit parametrization of $G_{\textrm{eff}}(z)$ was shown to have two important problems:
\begin{enumerate}
  \item It is a decreasing function of the redshift and therefore according to a general rule, the validity of which we demonstrated, it cannot be supported by a self-consistent scalar-tensor theory because it leads to a negative scalar field kinetic term.
  \item It predicts a large ISW effect that is not consistent with the observed large scale (low-$\ell$) CMB spectrum.
\end{enumerate}

These problems could potentially be resolved by considering more general modified gravity models which can potentially support the derived best fit $G_{\textrm{eff}}(z)$ such as Horndeski models \cite{Alonso:2016suf,Perenon:2015sla,Perenon:2016blf} or bimetric gravity\cite{Berg:2012kn}. The tension of the best fit $G_{\textrm{eff}}(z)$ with the low-$\ell$ CMB spectrum induced by the ISW effect is more difficult to resolve and may indicate either required modifications on the background \plcdm $H(z)$ or systematics in the growth data.

\begin{figure*}[!t]
\centering
\includegraphics[width = 0.98\textwidth]{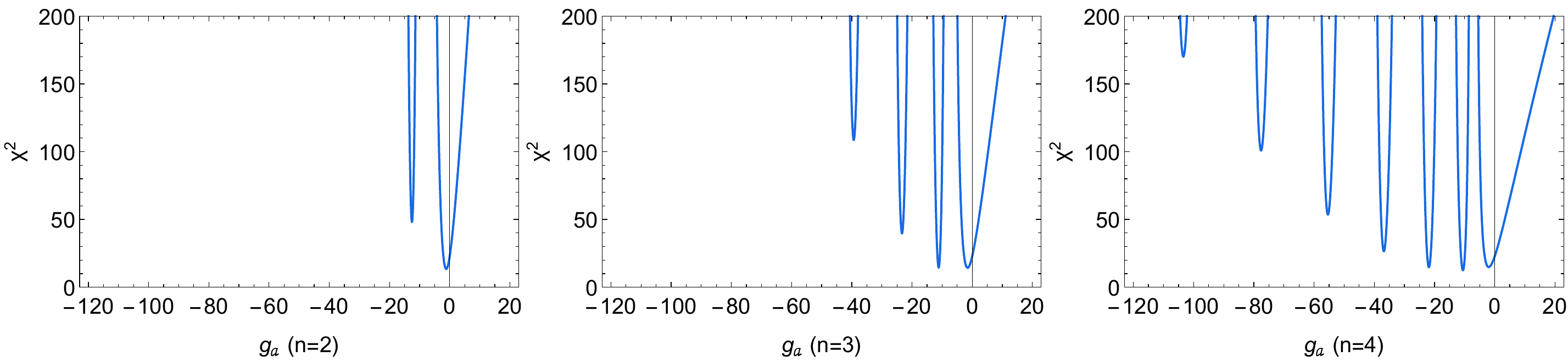}
\caption{Plots of the $\chi^2(g_a)$ for $n=2,3,4$ that clearly show the degeneracy of the model and the many minima of the $\chi^2$ in terms of $g_a$.}
\label{fig:minima}
\end{figure*}

The strategy of our analysis has been the identification of the consistency (or tension) of the \plcdm model with the growth data. In the context of this goal we have chosen to fix the \lcdm parameters to the Planck15 values. Clearly, the level of the tension can be reduced significantly if we vary the \lcdm parameters and in fact it may completely disappear if we consider background $H(z)$ parametrizations beyond \lcdm. However such an approach would not be consistent with the above described strategy.

We have pointed out the need for the construction of optimized, large, self-consistent compilations of the emerging growth data and have made a first attempt in that direction. Our updated `Gold-2017' dataset compilation comes from reliable sources, \ie major surveys and international collaborations. However, the fact that it consists of only a small amount of points indicates that there is significant potential for improvement. This situation will definitely improve in the coming decade as the Euclid \cite{Amendola:2016saw} and LSST \cite{Abell:2009aa} surveys will release a significant amount of new high quality data points and as a result, very soon we will be able to detect any possible deviations from GR with a high level of confidence.
\\ \\
\textbf{Numerical Analysis Files}: See Supplemental
Material at \href{http://leandros.physics.uoi.gr/growth-tension/}{here} for the Mathematica files used for the production of the figures, as well as the figures themselves.

\section*{Acknowledgements}
The authors thank Ms. Judit Perez for pointing out a couple of minor typos.

S.N. acknowledges support from the Research Project of the Spanish MINECO, FPA2013-47986-03-3P, the Centro de Excelencia Severo Ochoa Program SEV-2012-0249 and the Ram\'{o}n y Cajal programme through the grant RYC-2014-15843. IFT-UAM report number:
IFT-UAM/CSIC-17-031

\appendix
\section{Chaplygin theorem}
\label{sec:Appendix_A}
The Chaplygin theorem \cite{Chaplygintheorem} states that if $y(x)$ satisfies the $n$th-order differential inequality
\be
L[y]\equiv y^{n}(x)+a_1(x) y^{n-1}(x)+\cdots+ a_n(x)>b(x), \label{eq:difin1}
\ee
where the $a_n(x)$ can be integrated and the function $f(x)$ satisfies the differential equation
\be
L(f)=b(x)  \label{eq:difeq1}
\ee
with the same initial conditions as Eq.~(\ref{eq:difin1}), \ie $f(x_0)=y(x_0), \dots , f^{n-1}(x_0)=y^{n-1}(x_0)$, then there is a region $x\in(x_0,x_*]$ such that $y(x) > f(x)$ and $x_*$ is specified by the region for which for every $\xi \in[x_0,x]$ we have $G(s,\xi)\geq0$, where $G$ satisfies the Green equation with initial conditions
\ba
L[G]&=&0 \nn \\
G(x=\xi)&=&\cdots =G^{n-2}(x=\xi)=0, G^{n-1}(x=\xi)=1. \nn
\ea
By specifying the region where $G\geq0$ we can thus determine where $y(x)>f(x)$.

\section{Multiple minima}
\label{sec:Appendix_B}

A rather interesting feature that arises by minimizing the $\chi^2$ of the `Gold-2017' dataset using the $G_\text{eff}$ parametrization \eqref{geffansatz} is the one of the multiple minima. Specifically, as the number of $n$ increases the more minima we observe. This effect is due to the fact that the solution of the growth rate ODE of Eq.~(\ref{eq:ode}) contains Bessel functions which have degeneracies in their arguments. In order to keep things simple we will now consider a toy model with $\Omega_m=1$ and $G_{\textrm{eff}}/G_{\textrm{N}}=1+g_n(1-a)^n$, even though this model does not satisfy the viability criteria described in the text. Then, for $n=1$ and $\Omega_m=1$ the solution to the differential equation (\ref{eq:ode}) is
\ba
\delta_{n=1}(a)&=& c_1~a^{-1/4}~ J_{m}\left(\sqrt{6~a~g_n}\right)\nn \\
&+&c_2~a^{-1/4}~ J_{-m}\left(\sqrt{6~a~ g_n}\right),
\ea
while for $n=2$ we have
\ba
\delta_{n=2}(a)&=& e^{-\frac{1}{2} \beta a } a^{\frac{m}{2}-\frac{1}{4}} \left(c_1 U\left(\frac{1}{2} (m-\beta +1),m+1,a \beta \right) \right.\nn \\
&+&\left. c_2 L_{\frac{1}{2} (\beta -m-1)}^m (a \beta )\right),
\ea
where $c_{1,2}$ are constants to be determined for the growing and decaying mode and $m=\frac{1}{2} \sqrt{24 g_n+25}$, $\beta=\sqrt{6 g_n}$ and $J_{\pm m}\left( z \right)$, $U(\kappa_1,\kappa_2,z)$ and $L_n^\kappa$ are the BesselJ, the confluent hypergeometric $U$, and the Laguerre-L functions respectively. As can be seen in this case, the presence of these functions in the solution of the growth rate is the root cause of the multiple minima since the variable $g_n$ of the model appears both in the order and the argument of the functions. As a result, the growth will be degenerate with respect to $g_n$, \ie for many different values of $g_n$ we will have the same growth factor. Similar arguments can be made for any value of $\Omega_m$, since the case studied here ($\Omega_m=1$) is just a limit of the \lcdm model, but also for  $G_{\textrm{eff}}$ models, other than the one used in the main analysis.

Note that the first minimum is not always the global one, \ie the minimum with the smallest value of $\chi^2$. In Fig.~\ref{fig:minima} we show the $\chi^2(g_n)$ plot corresponding to our parametrization \eqref{geffansatz} for $n=2,3,4$, where for $n=2,3$ the first minimum is the global one, \eg for $n=3$ the first minimum corresponds to $(\chi^2=14.3, \ g_n=-1.534)$, while the second one corresponds to $(\chi^2=14.6, \ g_n=-11.14)$). On the other hand for $n=4$, the three first minima from right to left correspond to $(\chi^2 = 14.9, \ g_n = -2.006), (\chi^2 = 12.6, \ g_n = -10.56)$ and $(\chi^2 = 14.8, \ g_n = -21.87)$ respectively and therefore the global minimum is the second one (albeit with a small difference). However, since $g_n$ must be small, we consider only the first minimum, which for small values of $n$ is the global one as well.

\raggedleft
\bibliography{bibliography}

\end{document}